\documentclass[iop,jphysa,superscriptaddress,longbibliography,
eprint,
]{revtex4-2}
\usepackage[utf8]{inputenc}
\usepackage[T1]{fontenc}
\usepackage{amsmath,amssymb,amsfonts,amsthm,mathtools,mleftright}
\usepackage{dsfont}
\usepackage{graphicx}
\usepackage[dvipsnames]{xcolor}
\usepackage{hyperref}
\hypersetup{colorlinks=true, linkcolor=BrickRed, urlcolor=blue!50!black, citecolor=blue!50!black}
\usepackage{cleveref}
\usepackage[version=4]{mhchem} 
\usepackage[noend,noline]{algorithm2e}

\newcommand\FUB{\affiliation{Freie Universität Berlin, Department of Mathematics and Computer Science}}
\newcommand\ZIB{\affiliation{Zuse Institute Berlin}}

\begin{document}

\title{Open reaction-diffusion systems: bridging probabilistic theory and simulations across scales}

\author{Mauricio J. del Razo}\email{delrazo@zib.de} 
\ZIB
\FUB

\author{Margarita Kostré}
\ZIB

\date{\today}

\begin{abstract}
	Reaction-diffusion processes are the foundational model for a diverse range of complex systems, ranging from biochemical reactions to social agent-based phenomena. The underlying dynamics of these systems occur at the individual particle/agent level, and in realistic applications, they often display interaction with their environment through energy or material exchange with a reservoir. This requires intricate mathematical considerations, especially in the case of material exchange since the varying number of particles/agents results in ``on-the-fly'' modification of the system dimension. In this work, we first overview the probabilistic description of reaction-diffusion processes at the particle level, which readily handles varying number of particles. We then extend this model to consistently incorporate interactions with macroscopic material reservoirs. Based on the resulting expressions, we bridge the probabilistic description with macroscopic concentration-based descriptions for linear and nonlinear reaction-diffusion systems, as well as for an archetypal open reaction-diffusion system. Using these mathematical bridges across scales, we finally develop numerical schemes for open reaction-diffusion systems, which we implement in two illustrative examples. This work establishes a methodological workflow to bridge particle-based probabilistic descriptions with macroscopic concentration-based descriptions of reaction-diffusion in open settings, laying the foundations for a multiscale theoretical framework upon which to construct theory and simulation schemes that are consistent across scales.
\end{abstract}

\maketitle

\section{Introduction}

Living cells constantly exchange material and energy with their environment; they consume chemical energy and dissipate heat. Namely, they operate in an open non-equilibrium setting. In terms of physical chemistry, every living system must be an open system —a closed system has no life \cite{qian2007phosphorylation}. At biological cell scales, biochemical reaction systems can be characterized as an interplay between the spatial transport (diffusion) of molecules and their chemical kinetics (reaction) \cite{murray2003book,winkelmann2020stochastic}. Thus, the mathematical modeling of biochemical reaction systems in living cells requires the formulation of reaction-diffusion processes in open settings \cite{qian2010chemical}. Beyond applications in biochemical systems, reaction-diffusion offers a mathematical baseline to model a diverse range of complex systems, including the spread of diseases \cite{britton2019epidemic,malysheva2022stochastic,winkelmann2021mathematical}, innovations \cite{djurdjevac2018human,zonker2023insights}, opinions \cite{helfmann2023modelling} and other agent-based social systems. In real-world applications, it is prevalent that the systems of interest are open and operate in non-equilibrium regimes.

The formal description of reaction-diffusion dynamics as a stochastic process at the particle level is cumbersome due to the need to handle changes to the system's dimension as the dynamics evolve. The diffusion of individual particles/molecules is governed by Brownian dynamics while reactions can change the number of particles/molecules of the system, resulting in an ``on the fly''  modification of the system's dimension/degrees of freedom. This greatly complicates formulating the governing equations as one cannot simply write the stochastic differential equations of the dynamical system. In the seminal works by Doi \cite{doi1976second,doi1976stochastic}, the author derived a master equation for reaction-diffusion systems that describes the evolution of the probability distribution dynamics of reaction-diffusion processes in the phase space. This formalism has been later formalized and unified in recent work \cite{del2021probabilistic, delRazo2,del2024field, del2025dynamics} and \cite{isaacson2021reaction, isaacson2022mean}. This can be understood as follows: instead of focusing on the non-linear dynamics of individual particles, the dynamics are lifted to an infinite-dimensional phase space with a very peculiar structure ---symmetric with respect to particle permutations of the same type, continuous regarding particle's positions and discrete in particle number. In this space, the dynamics of the evolution of the probability distribution are conveniently linear. This is analogous to the Fokker-Planck equations for stochastic dynamical systems with a constant number of particles, with the main difference that this Master equation can handle a varying number of particles \cite{del2025dynamics} and thus can handle modeling reaction and diffusion jointly as one stochastic process. We refer to this master equation as the chemical diffusion master equation (CDME). 

The first work to present the CDME \cite{doi1976second} derives a field-theory inspired framework that can facilitate writing the equation and performing calculations. This framework is later employed to obtain path integral formulations \cite{peliti1985path,tauber2005applications,weber2017master}, which assume the rate functions are homogeneous in space. This is not the case in our current work; however, there is still a close connection with field theory inspired representations and path integrals of reaction-diffusion processes, as investigated in the recent review \cite{del2024field}. In this work, we do not discuss these aspects, and we focus on classical probabilistic representations to make the work accessible to a broader range of practitioners.

The main focus of this work is to investigate reaction-diffusion dynamics in an open systems context. In general, an open system refers to a system that exchanges matter and energy with an external system or reservoir \cite{qian2010chemical}. Energy exchanges are often easier to model since they don't change the dimension of the system and can often be modeled by incorporating some form of thermostat into the diffusion. However, material exchanges involve adding or removing particle/molecules/agents to the system and thus require changing the dimension of the system. In reaction events, the exchange of energy and matter with an external system is implicitly modeled through the rate function. For instance, reactions such as spontaneous creation $\emptyset \rightarrow A$ implicitly assume the particles are coming from somewhere or created given some energetic input, otherwise they would violate conservation laws. Moreover, spatially inhomogeneous systems in contact with a reservoir must incorporate particles/molecules in ways that are physically consistent with the diffusion process (or even hydrodynamics) across the contact boundary. To study and simulate open reaction-diffusion systems, we need to model energy and material exchange with reservoirs. As heat exchanges have been thoroughly studied in the literature \cite{hijon2010mori,zwanzig2001nonequilibrium}, we mainly focus on reservoir interactions that change the number of particles in the system through reactions or contact boundaries, while maintaining spatial resolution.

In this manuscript, we first overview the CDME and show how reactions are coupled to the diffusion process; we point the reader to the source works for more details. Then, starting from the CDME, we demonstrate how to incorporate interactions with a macroscopic reservoir while maintaining physical consistency. This is achieved by modeling diffusion exchanges across the contact boundary with the reservoir as ``reactions'' with specific characteristics. Further on, we introduce an intuitive methodology to recover macroscopic concentration-based descriptions from particle-based probabilistic descriptions and showcase this for three examples: linear and nonlinear reactions, as well as for an open reaction-diffusion system. Although the first two examples have been studied in the literature (e.g \cite{doi1976second,isaacson2021reaction,del2024field}), our approach is different and perhaps more intuitive for practitioners. Moreover, they are required for the last example, in which starting from the formulation of the CDME with reservoir interactions, we use statistical limits to yield a landmark example of macroscopic reaction-diffusion systems in an open setting (diffusion-influenced reaction models). Finally, we use these analytical results to develop numerical schemes and implement them into two illustrative simulations: an open reaction-diffusion system in contact with a reservoir and a reactive boundary, as well as a hybrid particle/concentration-based simulation of a SIR (Susceptible, Infectious, or Recovered) epidemiology model. This illustrates how the CDME framework can be used to establish mathematical relations between the models at different scales, enabling the development of multiscale and hybrid numerical schemes and thus bridging theory and simulation of open reaction-diffusion processes across scales.

In what follows, we present the CDME overview in \cref{sec:CDME}, along with two representative examples. In \cref{sec:openbnd}, we incorporate interactions between particle-based systems and macroscopic reservoirs through a contact boundary. \Cref{sec:BridgingCDME} focuses on recovering concentration-based descriptions of reaction-diffusion systems as a limiting case of the CDME for three representative examples: creation and degradation (linear), mutual annihilation (nonlinear) and diffusion-influenced reaction systems (linear, open). The last one is the main result in the context of open systems. \Cref{sec:resinteraction} uses the analytical results from the previous section to develop numerical schemes for particle-based reaction-diffusion system with reservoir interaction. We finalize with \cref{sec:simulations}, where we implement the numerical methods for two relevant examples.

\section{Probabilistic theory for reaction-diffusion processes: the chemical diffusion master equation}
\label{sec:CDME}

\begin{figure}[bt]
	\centering
	\includegraphics[width=0.8\textwidth]{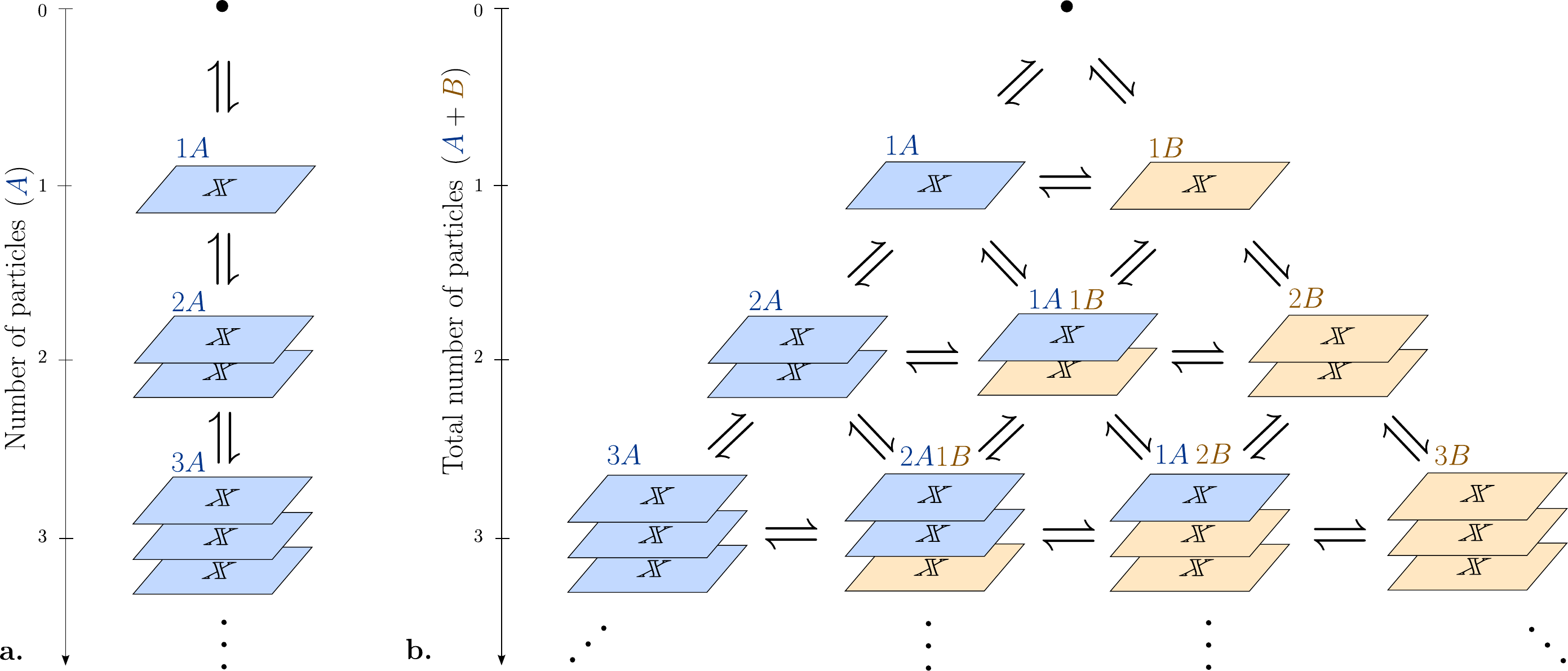}
	\caption{Structure of the phase space of the CDME. \textbf{a.} Phase space for a system with one chemical species $A$, consisting of discrete sets of continuous diffusion domains $\mathbb{X}$ depending on the number of particles. The transitions between sets depend on the specific reaction system. The arrows only show possible transitions between first neighbors. \textbf{b.} Analogous phase space but for a system with two chemical species $A$ and $B$. As particles of the same species are statistically indistinguishable, the ordering in phase space is not relevant.
    }
	\label{fig:phase+space_multi}
\end{figure} 

In this section, we overview the chemical diffusion master equation (CDME) \cite{del2021probabilistic,delRazo2, doi1976second}. This equation describes the probabilistic dynamics of reaction-diffusion processes at the particle level. As reactions can change the number of particles in the system, the system's degrees of freedom change with time. Thus, one cannot express the dynamics in terms of stochastic differential equations, and one has to instead describe the dynamics in terms of evolution of the probability distribution of the system, i.e. in terms of a master equation (or Kolmogorov forward equation) given in this case by the CDME. This distribution lives in a phase space with a complex structure that combines continuous and discrete degrees of freedom, see \cref{fig:phase+space_multi}. 

We consider a system of a varying number of particles of the same chemical species in a finite space domain $\mathbb{X}$ (e.g. a region within $\mathbb{R}^3$). The diffusion process changes the spatial configuration of the particles (continuous) while the reaction process can change the number of particles (discrete). The configuration of the system is given by the number of particles and their positions, and its probability distribution is given as a family/hierarchy of densities
\begin{align}
	\rho = \left( \rho_0, \rho_1(x^{(1)}), \rho_2(x^{(2)}), \dots, \rho_n(x^{(n)}), \dots \right),
\end{align}
where $\rho_n(x^{(n)})$ is the probability density of finding $n$ particles at positions $x^{(n)}=(x_1^{(n)},\dots, x_n^{(n)})$. We further assume each of these densities is symmetric with respect to particle permutations since they are statistically indistinguishable from each other. The normalization condition is 
\begin{align}
	\rho_0 + \sum_{n=1}^\infty \int_{\mathbb{X}^n} \rho_n(x^{(n)})dx^{(n)} = 1
	\label{eq:CDMEnormalization}
\end{align}
In general, $\rho$ will depend on time, but for simplicity, we omit time from the notation. As a remark, the density $\rho$ lives in a function space called Fock space \cite{del2021probabilistic, doi1976second, grassberger1980fock}. For $m$ reactions, the CDME then has the general form

\begin{align}
	\partial_t \rho = \left(\mathcal{D} + \sum_{r=1}^m\mathcal{R}^{(r)} \right)\rho,
	\label{eq:CDMEgeneral}
\end{align}
with $\mathcal{D}$ the diffusion operator and $\mathcal{R}^{(r)}$ the reaction operator corresponding to the $r$th reaction. It is assumed that diffusion is isotropic and without drift, so the $n$th component of the diffusion term corresponds to standard Brownian motion,
$\mathcal{D}_n \rho_n = \sum_{i}^n D_i \rho_n$, where $D_i=D\nabla^2_i$ with a scalar diffusion constant $D > 0$ and $\nabla^2_i$ acts on the $i$th component. Moreover, we impose a reflective boundary condition on $\mathbb{X}$ to the diffusion process (to each $\rho_n$). Note one can also write an analogous equation for the fluctuating concentration fields in terms of measure-valued processes \cite{isaacson2022mean,oelschlager1989derivation}. Under physically reasonable assumptions, these approaches are equivalent \cite{lim2020quantitative}.

\begin{figure}[h]
	\centering
	\includegraphics[width=0.45\textwidth]{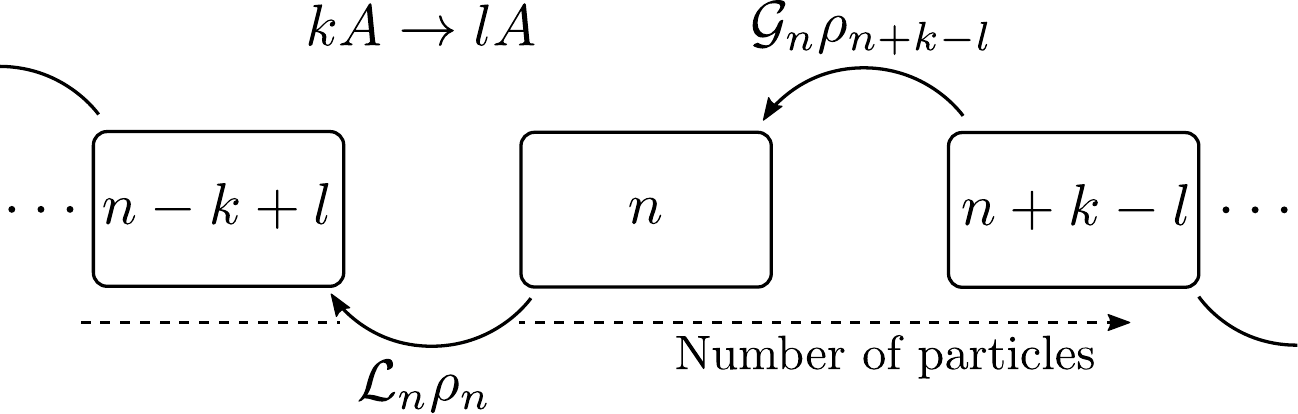}
	\caption{Diagram representing loss from and gain into the $n$ particle state due to the reaction $kA\rightarrow lA$ assuming $k>l$. The loss $\mathcal{L}_n \rho_n$ must depend on the reactant's positions, so it is a function of $x^{(n)}$, while the gain $\mathcal{G}_n\rho_{n+k-l}$ depends on the positions of the products, and thus must also be a function of $x^{(n)}$.}
	\label{fig:cdmstylediag}
\end{figure}

The reaction operator $\mathcal{R}^{(r)}$ corresponds to the $r$th reaction, and it is conveniently split into loss and gain operators, 
\begin{align*}
	\mathcal{R}^{(r)} = \mathcal{G}^{(r)} - \mathcal{L}^{(r)}, 
\end{align*}
each yielding the probability loss or gain to the current state due to the corresponding reaction. We can also write the CDME component-wise, e.g. for a system with one general one species reaction $kA\rightarrow lA$, the CDME is (\cref{fig:cdmstylediag})
\begin{align}
	\partial_t \rho_n = \mathcal{D}_n\rho_n + \mathcal{G}_n\rho_{n+k-l} - \mathcal{L}_n \rho_n \ ,
\end{align}
where $\mathcal{L}_n\rho_n$ yields the probability loss of the $n$-particle configuration into the $n-k+l$ one. Analogously, $\mathcal{G}_n\rho_{n+k-l}$ yields the probability gain into the $n$-particle configuration due to a reaction happening at the $n+k-l$ one, as shown in \cref{fig:cdmstylediag}. Conservation of probability establishes that the total probability loss of one state must correspond to the total probability gain of the receiving state, i.e. $\int \mathcal{L}_{m} \rho_{m} dx^{(m)} = \int \mathcal{G}_{m-k+l}\rho_{m}dx^{(m-k+l)}$ for any valid $m$. This was proved in \cite{del2021probabilistic}. This will be further clarified in the examples below.

If the diffusion averages out, the spatial dependence is lost, and one would recover the well-known chemical master equation \cite{qian2010chemical, winkelmann2020stochastic}. Alternatively, if there are no reactions, each component of the CDME simplifies to a Fokker-Planck equation for the diffusion of a fixed number of particles. For a more detailed explanation of how to derive the CDME for general systems, the reader is referred to \cite{delRazo2} and to \cite{del2021probabilistic} for a detailed account of the mathematical details of this equation. Below we formulate the equation at once for a couple of relevant examples.

\subsection{Degradation and creation}
\label{sec:degcreaCDME}
\begin{figure}[bt]
	\centering
	\includegraphics[width=0.9\textwidth]{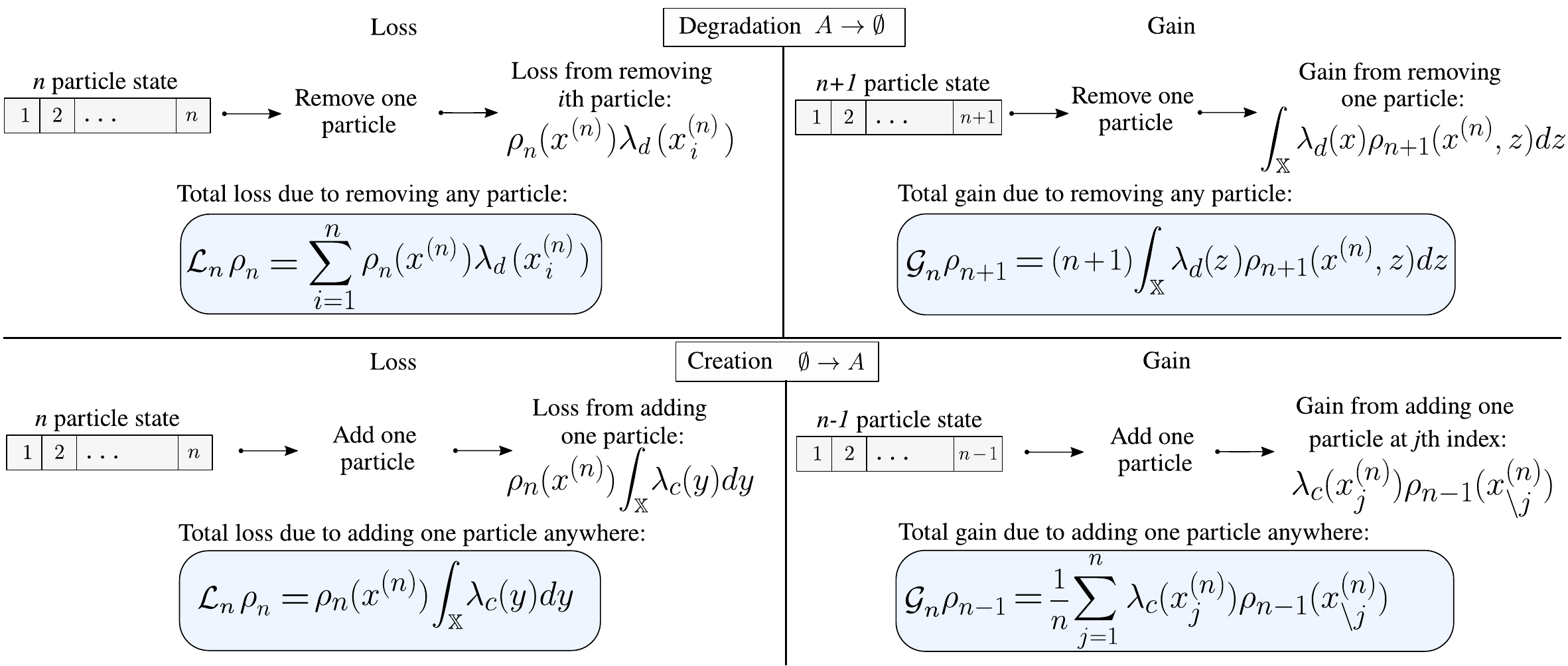}
	\caption{Schematic calculation of how to obtain the loss and gain terms of the CDME for degradation and creation reactions for the $n$-particle configuration. In all cases, it follows the same procedure: 1. calculate the loss or gain due to one possible reaction in terms of the rate function. 2. calculate the total loss due to any possible reaction. Note the rate function for the degradation case is a function of the reactant's position, while for the creation is a function of the product's position. In general, regardless of the reaction, both the loss and gain are functions of the particle configuration of the current state $x^{(n)}$. For the the loss, this dependence manifests in terms of the reactant's positions, while the product's positions are integrated out (if any). Conversely, for the gain, this manifests as dependence in terms of the product's positions, while the reactant's positions are integrated out(if any).}
	\label{fig:cdme_crea_deg}
\end{figure}

To derive the loss and gain terms of the CDME for degradation and creation reactions, we follow the schematic presented in \cref{fig:cdme_crea_deg}.
For a simple degradation reaction, $A\rightarrow \emptyset$, with rate function $\lambda_d(x)$, where $x$ is the position of the reactant, the $n$th component of the CDME has the following form (also see \cite{delRazo2, del2021probabilistic}),
\begin{align}
	\partial_t\rho_n(x^{(n)}) = \sum_{i=1}^n D_i \rho_n(x^{(n)}) - \sum_{i=1}^{n}   \lambda_d(x^{(n)}_i)\rho_n(x^{(n)}) + (n+1)\int_{\mathbb{X}} \lambda_d(y)\rho_{n+1}(x^{(n)},y) dy,
	\label{eq:CDME_degrad}
\end{align}
where $D_i=D\nabla^2_i$, the second term corresponds to the loss and the third to the gain. Also following \cref{fig:cdme_crea_deg}, for a creation reaction $\emptyset \rightarrow A$ with rate function $\lambda_c(x)$, where $x$ is the positions of the product, the CDME is given by
\begin{align}
	\partial_t\rho_n(x^{(n)}) &= \sum_{i=1}^n D_i \rho_n(x^{(n)}) - \rho_n(x^{(n)}) \int_{\mathbb{X}} \lambda_c(x) dx + \frac{1}{n}\sum_{i=1}^{n}\rho_{n-1}(x^{(n)}_{ \setminus i})  \lambda_c(x^{(n)}_i),
	\label{eq:CDME_creat}
\end{align}
where the subscript $\setminus i$ means that the entry with index $i$ is excluded from the tuple $x^{(n)}$ of particle positions.

\subsection{Bimolecular reactions} 
\label{sec:cdme_bimol}
Consider the reaction 
\[
A+B \rightarrow C
\]
with rate function $\lambda(z; x,y)$, where $x$ and $y$ are the locations of one pair of reactants and $z$ is the location of the product (note this rate function encodes the probability density of where the product particle is placed, see \cref{sec:secordreactSims}). The dynamics are described in terms of the distributions $\rho_{n,m,l}\left(x^{(n)};y^{(m)};z^{(l)}\right)$, where $n,m,l$ indicate the number of $A$, $B$ and $C$ particles and $x^{(n)}$, $y^{(m)}$ and $z^{(l)}$ their positions, respectively. The normalization condition is
\begin{align*}
	\sum_{n,m,l=0}^\infty \int_{\mathbb{X}^{(n+m+l)}} \rho_{n,m,l}\left(x^{(n)};x^{(m)};x^{(l)}\right) dx^{(n)}dx^{(m)} dx^{(l)} = 1.
\end{align*}

The CDME for this reaction has the same structure as before, and its derivation follows a similar logic to \cref{fig:cdme_crea_deg}, but with the additional complexity of handling multiple species and bimolecular reactions. The CDME for this reaction has been derived in previous works \cite{delRazo2,doi1976second,isaacson2013convergent,isaacson2018unstructured}; the $n,m,l$th component has the following form,
\begin{align}
	\begin{split}
		\partial_t \rho_{n,m,l} &= \sum_{i=1}^n D_{i}^A\rho_{n,m,l} + \sum_{j=1}^m D_{j}^B\rho_{n,m,l} +\sum_{k=1}^l D_{k}^C\rho_{n,m,l}  \\
		&- \sum_{i=1}^n \sum_{j=1}^m \rho_{n,m,l}\left(x^{(n)};y^{(m)};z^{(l)}\right) \int_{\mathbb{X}} \lambda\left(z'; x_{i}^{(n)},y_{j}^{(m)}\right) dz' \\
		& +\frac{(n+1)(m+1)}{l}\sum_{k=1}^{l}  \int_{\mathbb{X}^2} \rho_{n+1,m+1,l-1} 
		\left(x^{(n)},x';y^{(m)},y';z^{(l)}_{\setminus k} \right)
		\lambda\left(z^{(l)}_k; x',y'\right)dx'dy',
	\end{split}
	\label{eq:CDMEbim}
\end{align}
where to simplify notation, we skipped the explicit dependence of $\rho_{n,m,l}$ on positions and time if clear from the context. The operators $D_{i}^A$, $D_{i}^B$ and $D_{i}^C$ correspond to the one particle diffusion operators acting on the $i$th particle of the corresponding species. The second line corresponds to the loss and the third one to the gain. A simple approach to derive the CDME at once for more complex reaction-diffusion systems is to employ creation and annihilation operators in Fock space as first introduced by \cite{doi1976second} and further investigated in \cite{delRazo2,del2021probabilistic,del2024field}.

\section{Open boundaries and reservoirs at the particle level}
\label{sec:openbnd}

The mathematical formulation of the CDME can readily handle a variable number of particles. In this section, we show how to incorporate boundaries in contact with a reservoir into the CDME. This mathematical formal treatment is one of the novel contributions of this work. For simplicity, consider a system of particles of one chemical species diffusing with diffusion coefficient $D$ in a finite one-dimensional domain $\mathbb{X}$. The right end of the domain ($x=R$) is an open boundary in contact with the reservoir, which has a constant concentration $c_R$, while the left end ($x=0$), is closed (reflective).  To model the diffusion of incoming and outgoing particles at the open boundary, we need to add or remove particles from the system in the correct amounts, thus we can model them as creation/degradation reactions with a physically consistent choice of rate functions.

\begin{figure}[bt]
	\centering
	\textbf{a.}
	\includegraphics[width=0.35\textwidth]{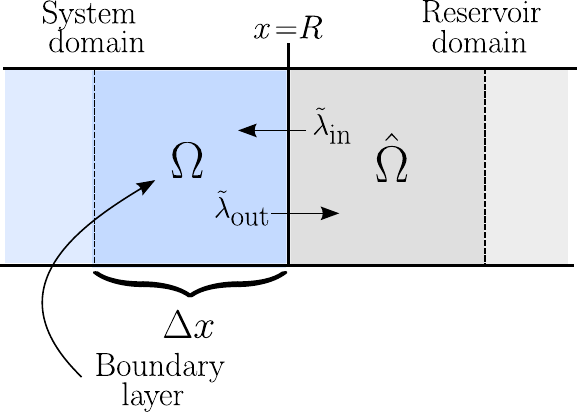}
	\qquad \qquad
	\textbf{b.} \includegraphics[width=0.35\textwidth]{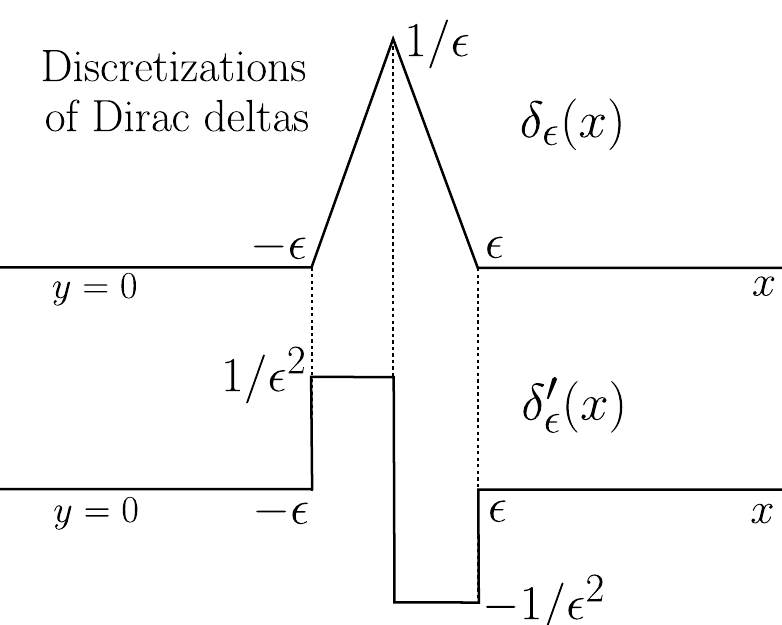}
	\caption{Discretizations diagrams. \textbf{a.} Illustration of the interaction between the system and the reservoir assuming the use of discrete rates from \cref{eq:generalRates}. The boundary layer $\Omega$ is shown as well as its mirror region in the reservoir domain $\hat{\Omega}$. \textbf{b.} Diagram of one possible discretization of the Dirac delta and its derivative as a function of a small parameter $\epsilon$. The discretization of the Dirac delta is denoted as $\delta_\epsilon(x)$, and its discretized derivative as $\delta'_\epsilon(x)$.
	}
	\label{fig:reservoir_interaction}
\end{figure}

To model the interaction with the reservoir, we draw inspiration from numerical approaches; we imagine a boundary layer glued at the right boundary. Particles move from the reservoir into the boundary layer with rate $\tilde{\lambda}_\text{in}$, and each particle can also go from the boundary layer into the reservoir with rate $\tilde{\lambda}_\text{out}$, see \cref{fig:reservoir_interaction}a. To obtain explicit expressions, consider a boundary layer $\Omega$ of length $\Delta x$, i.e. $\Omega = \{x | R-\Delta x \leq x \leq R\}$. Each particle in $\Omega$ can go into the reservoir by doing an outward ``diffusive jump''. As studied in \cite{del2018grand,kostre2021coupling,del2016discrete}, this corresponds to $D/\Delta x^2$, so 
\begin{equation}
	\tilde{\lambda}_\text{out}(x)= \frac{D}{\Delta x^2} \mathds{1}_{x\in\Omega},
	\label{eq:lout_discrete}
\end{equation}
where $\mathds{1}_{x\in \Omega}$ is the indicator function, giving a value of $1$ if $x\in \Omega$ and $0$ otherwise. We denote this rate with a tilde because it can be understood as a discrete approximation to a ``continuous'' rate acting on an infinitesimally small boundary layer. This expression hints that this rate should be given in terms of a ``derivative'' of a Dirac delta function (in the distributional sense)
\begin{equation}
	\lambda_\text{out}(x)= D\delta'(x-R),
	\label{eq:lout_cont} 
\end{equation}
such that a discretization yields $\tilde{\lambda}_\text{out}(x)$. This can be shown by first doing a simple discretization of the Dirac delta and its derivative in terms of a small parameter $\epsilon$. These are denoted by $\delta_\epsilon(x)$ and $\delta_\epsilon'(x)$ respectively
\begin{align}
	\delta_\epsilon(x)=
	\begin{cases}
		\displaystyle
		\frac{\epsilon-|x|}{\epsilon^2} \quad \text{if} \quad |x| <\epsilon \\[2mm]
		0 \qquad \ \ \text{otherwise,}
	\end{cases}
	\qquad \qquad
	\delta_\epsilon'(x)=
	\begin{cases}
		\displaystyle
		-\frac{1}{\epsilon^2}\text{sgn} (x) \quad \text{if} \quad |x| <\epsilon  \\[2mm]
		0 \qquad \ \ \ \ \ \text{otherwise.}
	\end{cases}
	\label{eq:discretedeltas}
\end{align}
The plots of these discretizations are shown in \cref{fig:reservoir_interaction}b. If we apply this discretization to $\delta'(x-R)$ with $\epsilon=\Delta x$ and centered at $R$, the only contribution to the system domain $(0,R)$ is the positive part. This can be written as $\mathds{1}_{x\in\Omega}/\Delta x^2$, which matches \cref{eq:lout_discrete}. Alternative discretizations are also possible, which emphasize the utility of writing it in the form of \cref{eq:lout_cont}.

Analogously, one can imagine particles within a mirrored region of the boundary layer, $\hat{\Omega}=\{x|R\leq x \leq R+\Delta x\}$, can jump from the reservoir into the boundary layer, each with the diffusive jump rate $D/\Delta x^2$. As the reservoir concentration is $c_R$, the number of particles within the mirrored region of the boundary layer (in the reservoir) will be $n_R= V_{\hat{\Omega}} c_R$, where $V_{\hat{\Omega}}$ is the volume of $\tilde{\Omega}$. In the one-dimensional case, the volume $V_{\hat{\Omega}}$ is $\Delta x$. The total rate of incoming particles from the reservoir into the boundary layer is the sum of all the individual jump rates, thus
\begin{equation}
	\tilde{\lambda}_\text{in}(x) = \frac{1}{V_{\hat{\Omega}}}\sum_{i=1}^{n_R}\frac{D}{\Delta x^2} \mathds{1}_{x\in\Omega}=  \frac{D}{\Delta x^2} c_R \mathds{1}_{x\in\Omega},
	\label{eq:lin_discrete}
\end{equation} 
where we need to divide by $V_{\hat{\Omega}}$ to get the right scaling with volume changes of the boundary layer. Analogously to the previous case, $\tilde{\lambda}_\text{in}(x)$ can be understood as a discretization of a ``continuous'' rate of incoming particles
\begin{equation}
	\lambda_\text{in}(x)= c_R D \delta'(x-R).
	\label{eq:lin_cont}
\end{equation}

Following the CDME expressions for creation and degradation reactions (\cref{eq:CDME_degrad,eq:CDME_creat}), we can substitute the rates we just derived and write the CDME for one diffusive species with an open boundary in contact with a reservoir
\begin{align}
	\partial_t\rho_n = \sum_{i=1}^n D_i \rho_n -& \sum_{i=1}^{n}   \lambda_\text{out}(x^{(n)}_i)\rho_n(x^{(n)}) + (n+1)\int_{\mathbb{X}} \lambda_\text{out}(y)\rho_{n+1}(x^{(n)},y) dy \notag \\
	&- \rho_n(x^{(n)}) \int_{\mathbb{X}} \lambda_\text{in}(x) dx + \frac{1}{n}\sum_{j=1}^{n}\rho_{n-1}(x^{(n)}_{\setminus j})  \lambda_\text{in}(x^{(n)}_j).
	\label{eq:CDMEreservoir}
\end{align}
In principle, as $\lambda_\text{out}$ is modeling diffusion into the reservoir, it should be possible to remove the term with $\lambda_\text{out}$ and change the boundary condition for the diffusion process to be absorbing on the corresponding boundary. However, we don't cover this case in detail. 
Although the ``continuous'' rates are more adequate to perform calculations with the CDME, the discrete rates will often be more useful to perform numerical simulations. In general, if the individual diffusive particle jumping rates are $\gamma_\text{out}$ and $\gamma_\text{in}$, the discrete rates will be given by
\begin{align}
	\tilde{\lambda}_\text{out}(x)= \gamma_\text{out} \mathds{1}_{x\in\Omega}
	\qquad \qquad
	\tilde{\lambda}_\text{in}(x) = \gamma_\text{in} c_R \mathds{1}_{x\in\Omega} \qquad
	\label{eq:generalRates}
\end{align} 

These discrete rates are valid for discretizations in rectangular grids. In particular, if the diffusion process is being approximated by a random walk in a lattice, it has been shown in \cite{del2016discrete,kostre2021coupling} that there is a relation between the random walk timestep $\Delta t$ and the width of the boundary layer: $\Delta t \leq \Delta x^2/(2D)$. This is analogous to the Courant–Friedrichs–Lewy (CFL) condition from computational fluid dynamics. However, if the geometry of the boundary with the reservoir is curved, this should be taken into account in the discretization. The beauty of having these expressions in terms of Dirac deltas is not only that we can use them for analytical calculation, but we can also discretize them for different geometries using standard techniques \cite{peskin2002immersed,li2006immersed}. This is the main difference in comparison to the works \cite{del2016discrete,del2018grand,kostre2021coupling}; by establishing a direct connection with the underlying continuous probabilisitc dynamics given by the CDME, we enable a framework to systematically develop consistent and convergent numerical schemes based on the ground model.

\subsection{Handling reactions at the reservoir interface}
\label{sec:reactInterface}
Zeroth and first-order reaction depend on zero or one particle, and thus they do not require any additional consideration at the reservoir interface. However, second-order reactions depend on two particles, so a reactant in the particle domain can react with another reactant in the reservoir. Thus, if the particle type from the reservoir is involved in a second-order reaction, we need to take these reactions into account, otherwise, we risk losing accuracy in the boundary region and consequently in the entire domain.

One can incorporate these effects by adding a term into the CDME. Consider a system with $A$ and $B$ particles in contact with a reservoir of $B$ particles with concentration $c_R$. The particles react following $A+B\rightarrow \emptyset$ with $\lambda(x,y)$, where $x$ is the position of an $A$ particle and $y$ of a $B$ particle. To emulate implicitly a reaction between an $A$ particle in the system with a $B$ particle from the reservoir, we add a reaction $A\rightarrow \emptyset$ with rate $\hat{\lambda}(x)$ given by
\begin{align}
	\hat{\lambda}(x) = c_R\int_\mathcal{R} \lambda(x,y) dy,
	\label{eq:ratefuncAtInterface}
\end{align}	
where $\mathcal{R}$ is the reservoir domain. This implicitly models the reactions with particles in the reservoir without explicitly modeling them. Following \cref{sec:cdme_bimol} and \cite{delRazo2}, the resulting CDME will be
\begin{align}
	\begin{split}
	\frac{\partial \rho_{n,m}}{\partial t} &= \sum_{i=1}^n D_{A_i}\rho_{n,m} + \sum_{j=1}^m D_{B_j}\rho_{n,m}  
	- \sum_{i=1}^n \sum_{j=1}^m \rho_{n,m}\left(x^{(n)};y^{(m)}\right) \lambda\left(x_{i}^{(n)},y_{j}^{(m)}\right) \\
	& +(n+1)(m+1) \int_{\mathbb{X}^2} \rho_{n+1,m+1} 
	\left(x^{(n)},x';y^{(m)},y' \right)
	\lambda\left(x',y'\right)dx'dy', \\
	&- \sum_{i=1}^{m}   \lambda_\text{out}(y^{(n)}_i)\rho_{n,m}\left(x^{(n)};y^{(m)}\right) + (m+1)\int_{\mathbb{X}} \lambda_\text{out}(y')\rho_{n+1}(x^{(n)},;y^{(m)},y') dy' \\
	&- \rho_{n,m}\left(x^{(n)};y^{(m)}\right) \int_{\mathbb{X}} \lambda_\text{in}(y) dy + \frac{1}{m}\sum_{j=1}^{m}\rho_{n,m-1}(x^{(n)};y^{(m)}_{\setminus j})  \lambda_\text{in}(y^{(m)}_j)\\
	& - \sum_{i=1}^{n}   \hat{\lambda}(x^{(n)}_i)\rho_{n,m}\left(x^{(n)};y^{(m)}\right) + (n+1)\int_{\mathbb{X}} \hat{\lambda}(y)\rho_{n+1,m}(x^{(n)},y) dy,
	\end{split}
	\label{eq:CDMEresreact}
\end{align}
where $\rho_{n,m}\left(x^{(n)};y^{(m)}\right)$ is the probability of having $n$ particles $A$ at positions $x^{(n)}$ and $m$ particles $B$ at positions $y^{(m)}$. The first two lines correspond to the standard CDME for $A+B\rightarrow \emptyset$ (see \cite{delRazo2})); the next two lines correspond to the interaction with a reservoir of $B$ particles as derived in \cref{sec:openbnd}; and the last line incorporates possible reactions across the boundary with the reservoir in the form of a degradation reaction with rate $\hat{\lambda}$. Although this equation seems cumbersome, it shows we can write the reactions occurring across the reservoir interface in terms of an alternate reaction with a specific rate function, facilitating the development of numerical schemes, as shown in \cref{sec:resinteraction}.

\section{Bridging reaction-diffusion across scales}
\label{sec:BridgingCDME}

The CDME is a probabilistic description of reaction-diffusion processes at the particle level, and it thus contains all the information of the underlying process required to recover coarser descriptions. It serves as a ground model to derive models at coarser scales. In principle, one can derive not only macroscopic and mesoscopic models from the CDME but also information on the stochastic fluctuations and their relations between parameters at multiple scales. 

As the scope of this paper is in open systems, we focus on bridging the CDME to macroscopic open reaction-diffusion systems. This will be fundamental to developing consistent multiscale simulations of particle systems coupled with reservoirs. The properties of reservoirs are often defined by macroscopic thermodynamic quantities, such as concentrations or temperature, thus we need to ensure that the particle interaction with the reservoir is consistent with the macroscopic description of the process.

\begin{figure}[bt]
\centering
\includegraphics[width=0.8\textwidth]{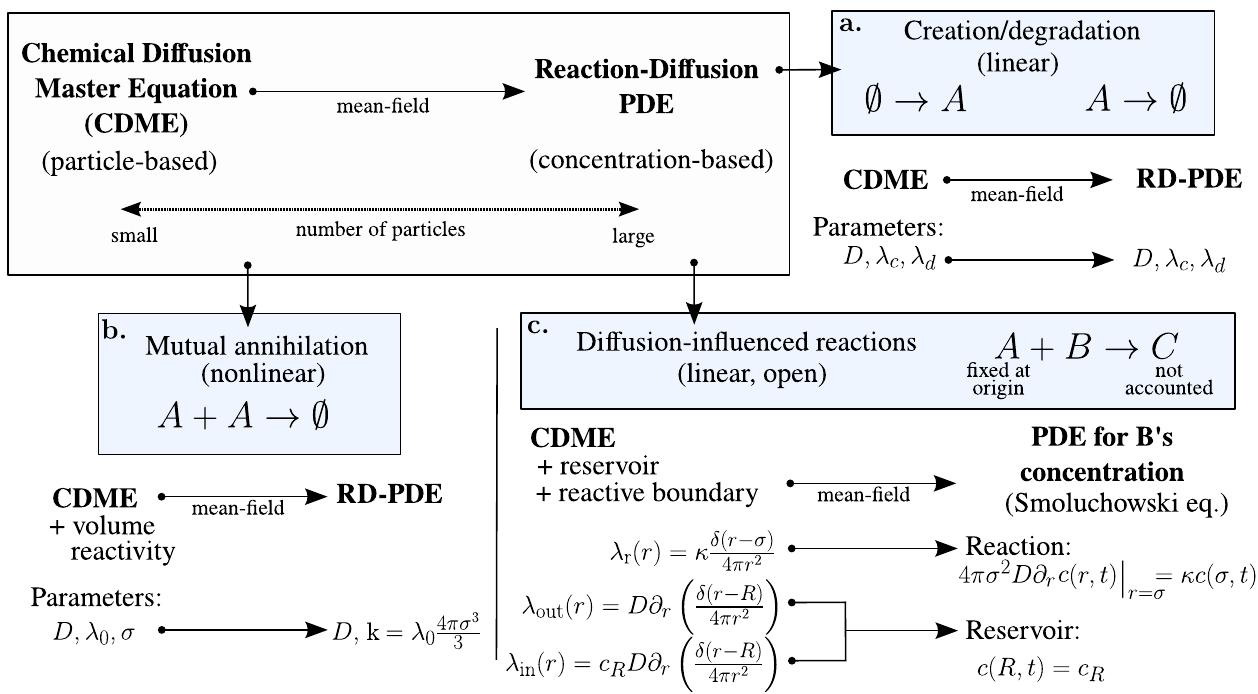}
\caption{Diagram of connections between the CDME and the reaction-diffusion PDEs (RD-PDEs) for three examples, as well as the connection between the parameters at the different scales. The PDE descriptions are recovered by taking the mean-field. Note that for linear systems the mean-field is equivalent to the large copy number limit \cite{del2018grand,kostre2021coupling}. \textbf{a.} For simple linear reactions, like creation and degradation, the parameters at the particle level match those at the concentration description. \textbf{b.} For nonlinear reactions involving two reactants, such as mutual annihilation and bimolecular reactions, we choose the volume reactivity model for the rate function \cite{doi1976stochastic}. This depends on a microscopic rate $\lambda_0$ and reactive distance $\sigma$. Our work yields the macroscopic rate in the RD-PDE in terms of the microscopic parameters. Note in this case one also requires to assume large copy numbers to neglect the covariance arising from the nonlinear reaction \cite{kostre2021coupling}. \textbf{c.} Although the reaction modeled is in principle nonlinear, the model focuses on the concentration of $B$ modulated by one reservoir on one edge and a reactive boundary on the other, so the particle dynamics are effectively linear, i.e. there are no pair interactions. Our approach shows how to consistently choose the microscopic rate functions to match the macroscopic description of the open system. 
}
\label{fig:diagsCDME-RDPDE}
\end{figure}

In this section, we present how to ---starting from the CDME--- recover a well-known class of physically relevant models of macroscopic reaction-diffusion in an open setting: diffusion-influenced reactions models. To recover this type of models, we first need to show how to recover the macroscopic reaction-diffusion equations for simple creation and degradation reactions. For the sake of completeness, we also show this last calculation for mutual annihilation ---the most simple nonlinear reaction--- before jumping into the main result of this section. Note the following results from \cref{sec:PDEdegreat,subsec:mutannih} have been proved rigorously in \cite{lim2020quantitative,isaacson2021reaction,isaacson2022mean,oelschlager1989derivation}. However, our approach is new, and we believe it is more intuitive and inline with methodologies used by chemical physicists. It further provides the basis for the novel result in \cref{sec:diffinflCDMElimit}. A summary of all these results is shown in \cref{fig:diagsCDME-RDPDE}.

\subsection{Degradation and creation} 
\label{sec:PDEdegreat}
The starting point of this section is the CDME for degradation and creation reactions \cref{eq:CDME_degrad,eq:CDME_creat}. The average concentration at point $y$, $c(y)$, can be recovered from the CDME for one species using the following formula in the \cref{sec:CDMEtoconc} and \cite{del2021probabilistic}:
\begin{align}
c(y) = \mathbb{E}[C(y)]=\sum_{n=1}^\infty n \int_{\mathbb{X}^{n-1}} \, \rho_n(y, x^{(n-1)}) \, dx^{(n-1)}.
\label{eq:conc_alt}
\end{align}
The quantity $C(y)$ corresponds to the stochastic concentration. This equation is obtained by calculating the expected number of particles in a ball of radius $\epsilon$ centered at $y$, dividing by the volume of the ball and taking the limit $\epsilon\rightarrow  0$. Applying this formula to \cref{eq:CDME_degrad,eq:CDME_creat}, respectively, we obtain the following macroscopic reaction-diffusion equations for the average concentration for the degradation and creation cases respectively
\begin{align}
\begin{split}
	\partial_t c(y,t) &= D \nabla^2 c(y,t) -\lambda_d(y) c(y,t), \\
	\partial_t c(y,t) &= D \nabla^2 c(y,t) +\lambda_c(y).
\end{split}
\end{align}
The calculation details are shown in \cref{sec:CDMEtoconc}. For higher-order reactions, these calculations become more involved and yield relations between the equation parameters at different scales. We show next this calculation for the simplest bimolecular case.

\subsection{Mutual annihilation}
\label{subsec:mutannih}
To simplify the presentation, instead of using the multispecies bimolecular reaction from \cref{sec:cdme_bimol}, we focus on the mutual annihilation reaction $A+A \rightarrow \emptyset$, which is the simplest since it only involves one chemical species. however, the result holds for the multispecies case as well. Assuming the rate function of the reaction is $\lambda(x_1,x_2)$ where $x_1$ and $x_2$ are the positions of the reactants, the CDME is derived analogously as before \cite{del2021probabilistic,delRazo2}
\begin{align}\begin{split} \label{eq:CDME_annihilation}
	\partial_t \rho_n(x^{(n)}) &=
	\mathcal{D}_n\rho_n(x^{(n)})  \quad + \frac{(n+1)(n+2)}{2} \int_{\mathbb{X}^2} \lambda(z_1,z_2)  \rho_{n+2}(z_1,z_2,x^{(n)}) \, dz_1 dz_2  \\
	& \quad  -\rho_n(x^{(n)}) \sum_{1\leq i <j \leq n}^n \lambda(x^{(n)}_i,x^{(n)}_j)   \,.
\end{split}
\end{align}
We can again apply formula \cref{eq:conc_alt} and following the steps in \cref{sec:CDMEtoconc}, we obtain the resulting equation:
\begin{align}
\partial_t c(y) = D \nabla^2 c(y) -\sum_{n=1}^\infty n (n-1) \int_{\mathbb{X}^{n-1}}  \rho_n(y,x^{(n-1)}) \lambda(y,x^{(n-1)}_1) dx^{(n-1)}.
\label{eq:RDMEpreDoi}
\end{align}
This equation is not closed since we still have the dependence on $\rho_n$. However, the term with the integral can be simplified further if we assume the reaction rate function has the form of an indicator function like in the well-known volume reactivity model by Doi\cite{doi1976stochastic}, i.e. 
\begin{align}
\lambda(y,z) = \lambda_0 \mathds{1}_{|y-z|\leq \sigma}.
\label{eq:doireactratefunc}
\end{align}
This means the reaction rate is $\lambda_0$ if particles are closer than a distance $\sigma$ or zero otherwise. In this case, the reaction term of \cref{eq:RDMEpreDoi} simplifies to
\begin{align}
-\lambda_0 \sum_{n=1}^\infty n (n-1)& \int_{|y-z|\leq \sigma}\int_{\mathbb{X}^{n-2}}  \rho_n(y,z,x^{(n-2)})  dx^{(n-2)}dz, 
= -\lambda_0 \int_{|y-z|\leq \sigma}\mathbb{E}[C(y)C(z)] dz,
\label{eq:reacterm_mutannih_integral}
\end{align}
where one recognizes the expectation of the product of concentrations, see \cref{eq:conccov}. This already allows us to write a PDE in terms of concentrations
\begin{align}
\partial_t c(y) = D \nabla^2 c(y) -\lambda_0 \int_{|y-z|\leq \sigma}\mathbb{E}[C(y)C(z)] dz
\label{eq:RDPDEmutannihcov}
\end{align}
An analogous expression to this one was obtained by Doi \cite{doi1976stochastic} and more rigorously analyzed in \cite{isaacson2022mean}. The expected value of the product of the stochastic concentrations at $y$ and $z$ can be rewritten as $\mathbb{E}[C(y)C(z)]=\mathbb{E}[C(y)] \mathbb{E}[C(z)] + \mathrm{cov}[C(y),C(z)]$. In the large volume and large copy number limit, the covariance is negligible in comparison to the product of the means \cite{kostre2021coupling}, so we can approximate $\mathbb{E}[C(y)C(z)]\approx \mathbb{E}[C(y)] \mathbb{E}[C(z)] =c(y)c(z)$, then the reaction term simplifies further, assuming we are in three dimensions 
\begin{align}
-\lambda_0\int_{|y-z|\leq \sigma}c(y)c(z)dz &\approx
-\lambda_0 c(y)^2 \int_{|y-z|\leq \sigma}dz \\
&=-\frac{4}{3}\pi \sigma^3 \lambda_0 c(y)^2  
\end{align}
where the approximation in the first line consists of assuming $\sigma$ is small enough, so the concentration $c(z)$ in $|y-z|\leq \sigma$ is well approximated by $c(y)$. The macroscopic rate is then given in terms of the microscopic parameters by the relation
\begin{align*}
k=  \frac{4}{3}\pi \sigma^3 \lambda_0 = V_\sigma \lambda_0.
\end{align*}
We recognize $V_\sigma$ is simply the volume of the reactive region $|y-z|\leq \sigma$, so this result remains true in any other dimension other than three. The reaction-diffusion PDE reincorporating time in the notation is then
\begin{align}
\partial_t c(y,t) = D \nabla^2 c(y,t) - k c(y,t)^2.
\label{eq:MFlimmutanih}
\end{align}
Note this result was first derived in \cite{doi1976stochastic} and later in \cite{isaacson2021reaction,kostre2021coupling}; though, here we employ a new and different method. Also note that to obtain this result we assumed the reaction rate function is an indicator function and with two approximations: covariance is zero and $c(z)\approx c(y)$ in the region $|x-z|\leq \sigma$. However, based on the integral term in \cref{eq:RDMEpreDoi}, we could make different choices of rate functions and approximations to derive alternative meso- and macroscopic models for the application at hand, while maintaining consistency across scales. The last and main result of this section is to show macroscopic diffusion-influenced reactions models can also be recovered from the CDME.

\subsection{Diffusion-influenced reactions as a limit of the CDME}
\label{sec:diffinflCDMElimit}

Diffusion-controlled/influenced reactions \cite{agmon1990theory,collins1949diffusion, doi1976stochastic, smoluchowski1918versuch, keizer1987diffusion, rice1985diffusion, shoup1982role} are archetypal models of a reaction-diffusion system in an open setting. Its original presentation---the Smoluchowski's model \cite{smoluchowski1918versuch, collins1949diffusion}--consists of a system with a chemical concentration gradient that is in contact with a constant concentration reservoir in the far-field and a reactive boundary close to the origin. These models have been fundamental in the development and understanding of reaction-rate theory \cite{hanggi1990reaction}, as well as in the development of experimental techniques (e.g. \cite{elson2011fluorescence,del2014fluorescence}). 

More specifically, they describe the reaction between a macromolecule $A$ and smaller molecules $B$: $A+B\rightarrow C$. The setup is as follows: assume there is one $A$ fixed at the origin and many $B$ molecules (described by a concentration gradient) diffuse freely in three dimensions around $A$ with a diffusion coefficient $D$. If a $B$ molecule reaches $A$, it can react with some given rate $\kappa$. On the far field, we assume there is a reservoir of $B$ molecules with constant concentration $c_R$, so the system is open. Given the symmetry of the problem, we can focus on the dynamics of chemical $B$ along the radial variable $r$. The concentration gradient of $B$ molecules around $A$ is denoted by $c(r,t)$, and it obeys the  Smoluchowski equation \cite{collins1949diffusion,smoluchowski1918versuch}
\begin{align}
\partial_t c(r,t) = D\nabla^2 c(r,t), \quad
\label{eq:smoleq}
\end{align}
where the Laplacian is in spherical coordinates and with boundary conditions
\begin{align}
4 \pi \sigma^2 D  \partial_r c(r, t) \Big|_{r=\sigma} = \kappa c(\sigma, t), \qquad
c(R, t) = c_R.
\label{eq:smoleqBC}
\end{align}
The reaction is modeled by the first boundary condition at the reaction boundary $r = \sigma$ (the sum of the molecules $A$ and $B$ radii). It states that the flux of $B$ particles across this boundary is proportional to the concentration at the boundary. At the particle level, this means that whenever a $B$ molecule reaches $\sigma$ by diffusion, a reaction occurs with rate $\kappa$, which controls the degree of diffusion influence in the reaction rate. The purely diffusion-controlled result $c(\sigma, t)=0$ is recovered as a special case in the limit  $\kappa \rightarrow \infty$ \cite{collins1949diffusion,smoluchowski1918versuch}. The second boundary condition models the system being in contact with the reservoir at $r = R$ with constant concentration $c_R$.

One can recast this problem probabilistically at the particle level by writing a CDME in an open setting. Assume an arbitrary and variable number of $B$ particles diffuse in a three-dimensional domain delimited by the spherical region $r\in[\sigma,R]$. We further assume the diffusion process itself is reflective at both boundaries $r=\sigma$ and $r=R$. However, if a $B$ particle reaches $r=\sigma$, a degradation reaction can occur with rate $\kappa$. In the far-field $r=R$, the system is in contact with a reservoir with constant concentration $c_R$ (also see \cref{fig:smol_convergence}). Following \cref{sec:openbnd}, this can be modeled as a combination of one creation and one degradation reaction (\cref{sec:degcreaCDME}). Thus, we can now write the CDME for this system
\begin{align}
\frac{\partial\rho_n}{\partial t} = D\sum_{i=1}^n \nabla^2_i \rho_n 
-&\sum_{i=1}^{n} \lambda_\text{r}(x^{(n)}_i)\rho_n(x^{(n)}) + (n+1)\int_{\mathbb{X}} \lambda_\text{r}(y)\rho_{n+1}(x^{(n)},y) dy
\notag \\
-& \sum_{i=1}^{n}   \lambda_\text{out}(x^{(n)}_i)\rho_n(x^{(n)}) + (n+1)\int_{\mathbb{X}} \lambda_\text{out}(y)\rho_{n+1}(x^{(n)},y) dy \notag \\
&- \rho_n(x^{(n)}) \int_{\mathbb{X}} \lambda_\text{in}(x) dx + \frac{1}{n}\sum_{j=1}^{n}\rho_{n-1}(x^{(n)}_{\setminus j})  \lambda_\text{in}(x^{(n)}_j),
\label{difflimiCDME}
\end{align}
with

\begin{align}
	\lambda_\text{r}(x) = \kappa\delta_\sigma(x), \qquad 
	\lambda_\text{out}(x) = D\nabla\delta_R(x), \qquad
	\lambda_\text{in}(x) = c_R D \nabla\delta_R(x),
	\label{eq:diffinflCDMErates0}
\end{align}
with $\delta_\sigma(x)$ a normalized surface delta function on the sphere of radius $\sigma$, i.e. $\delta_\sigma(x)=\delta(||x||-\sigma)/(4\pi ||x||^2)$.
The rate function  $\lambda_\text{r}$ models that particles reaching $r=\sigma$ react with rate $\kappa$.
The interaction with the reservoir is modulated by $\lambda_\text{out}$ and $\lambda_\text{in}$ as done in \cref{sec:openbnd}, where the derivatives were switched to gradients since now we are in three dimensions. This CDME is more general than the original Smoluchowski's model as it describes the evolution of the probability distribution. The original model from \cref{eq:smoleq,eq:smoleqBC} should be recovered in the mean-field limit.

As the interactions at the boundaries are regulated by the reaction and the reservoir, we can model them all in terms of creation or degradation reactions. Thus, to recover \cref{eq:smoleq}, we use the results from \cref{sec:PDEdegreat} for diffusion and creation/degradation reactions to obtain an equation for the average concentration 
\begin{align}
\partial_t c(x,t) = D\nabla^2 c(x,t) - \lambda_\text{r}(x) c(x,t) - \lambda_\text{out}(x) c(x,t) + \lambda_\text{in}(x).
\label{eq:difinfCDMElim}
\end{align}
The diffusion term is already analogous to the one in \cref{eq:smoleq}. We only need to show that the source terms of this equation are equivalent to the boundary conditions from \cref{eq:smoleqBC}. Due to the spherical symmetry of the problem neither the concentration nor the rates depend on the angular coordinates, so the equation in spherical coordinates is
\begin{align}
\partial_t \left(c(r,t)\right) = \frac{D}{r^2} \partial_r \left(r^2 \partial_r c(r,t)\right) - \lambda_\text{r}(r) c(r,t) - \lambda_\text{out}(r) c(r,t) + \lambda_\text{in}(r),	\label{eq:difinfCDMElim_sph}
\end{align}
with
\begin{align}
\lambda_\text{r}(r) = \kappa\frac{\delta(r-\sigma)}{4\pi r^2}, \qquad 
\lambda_\text{out}(r) = D\partial_r\left(\frac{\delta(r-R)}{4\pi r^2}\right), \qquad
\lambda_\text{in}(r) = c_R D \partial_r\left(\frac{\delta(r-R)}{4\pi r^2}\right).
\label{eq:diffinflCDMErates}
\end{align} 
To recover the first boundary condition at $r=\sigma$; we integrate \cref{eq:difinfCDMElim_sph} across the boundary from $\sigma - \epsilon$ to $\sigma + \epsilon$ in spherical coordinates, which yields
\begin{align}
\partial_t	\int_{\sigma - \epsilon}^{\sigma + \epsilon} c(r,t) 4\pi r^2 dr &= 
4\pi D \Big[r^2\partial_r c(r,t)\Big]_{\sigma-\epsilon}^{\sigma+\epsilon} - \int_{\sigma - \epsilon}^{\sigma + \epsilon}\kappa c(r,t)\delta(r-\sigma) dr, 
\end{align}
since all the other terms are zero close to $r=\sigma$. Due to continuity, as $\epsilon \rightarrow 0$, then
\begin{align}
4\pi \sigma^2 D  \partial_r (r,t)\Big|_{r=\sigma} &= \kappa c(\sigma,t),
\end{align}
where we assumed the concentration at $r<\sigma$ is zero and thus $\partial c/\partial r$ is also zero in this region. One can show this formally by solving the PDE in the domain $r\in[0,R]$ and use the jump of the derivative at $r=\sigma$ as an interface condition (also shown in \cite{zhang2022detailed}). This recovers the first boundary condition from \cref{eq:smoleqBC}.

To recover the second boundary condition,  we do not get useful information if we integrate across the
boundary $r=R$. Instead, we first do a spatial indefinite integral of \cref{eq:difinfCDMElim_sph} 
\begin{align}
\partial_t	\int c(r,t) 4\pi r^2 dr - 4\pi D r^2\frac{\partial c(r,t)}{\partial r} - \int \lambda_r c(r,t)4 \pi r^2dr + A
= D \int  \partial_r \left(\frac{\delta(x-R)}{r^2}\right)\left(c_R -c(r,t)  \right)  r^2dr,
\end{align}
with $A$ an integration constant. The integral on the right-hand side is solved by using integration by parts ($\int \delta'(x)f(x)dx = f(x)\delta(x)-f'(0)H(x)+\text{constant}$). Then, we integrate again from $R - \epsilon$ to $R + \epsilon$, and as we take the limit $\epsilon \rightarrow 0$, all the terms on the left-hand side as well as constants will vanish due to continuity. Thus, we are left with
\begin{align}
\lim_{\epsilon\rightarrow0}\int_{R-\epsilon}^{R+\epsilon}
\left[
\delta(x-R)\bigg(c_R-c(R,t)\bigg) - 
H(r-R)\left(\frac{2}{R}c_R-\frac{2}{R}c(R,t)-c'(R,t)\right)
\right]4\pi Dr^2dr 
=0
\end{align}

with $c'(r,t)=\partial_r c(r,t)$ and $H(x)$ the Heaviside function. The terms with the Heaviside function also vanish as $\epsilon \rightarrow 0$, and we are left with the integral with the Dirac delta, which implies
\begin{align}
c(R,t) = c_R,
\end{align}
yielding the second boundary condition from \cref{eq:smoleqBC}. We just showed \cref{eq:difinfCDMElim} is equivalent to the original formulation of diffusion-influenced reactions (\cref{eq:smoleq} with boundary conditions given by \cref{eq:smoleqBC}). Thus, the original formulation of diffusion-influenced reactions can be understood as the macroscopic limit that emerges when taking average concentrations of the corresponding CDME. It can also be shown this is equivalent to the large copy number limit \cite{del2018grand}.
Other formulations of diffusion-influenced reaction theory, including \cite{agmon1990theory,szabo1980first, szabo1989theory} can be reformulated as special cases of the CDME following similar procedures.

\subsubsection*{A remark about the interpretation of diffusion-influenced reactions models}
In its original formulation, the theory of diffusion-influenced reactions is applied to the system $A+B\rightarrow C$ to obtain an effective diffusion-influenced reaction rate. This is calculated as the flux at the reactive boundary $r=\sigma$ of the steady state solution to \cref{eq:smoleq,eq:smoleqBC}  as $R\rightarrow \infty$ \cite{del2016discrete}. This yields the diffusion-influenced reaction rate
\begin{align}
\hat{k}_\text{D}(\kappa) = \frac{\kappa k_S}{\kappa + k_S} \ , \qquad\text{with}\quad k_S=4\pi\sigma D,
\label{eqSmolsrates}
\end{align}
where in the limit $\kappa \rightarrow \infty$, $\hat{k}_\text{D}$ tends to the diffusion-controlled rate $k_S$ ---original Smoluchowski's result \cite{smoluchowski1918versuch}. An alternative approach is to substitute the reactive boundary condition from \cref{eq:smoleqBC}) by the volume reactivity model from \cref{eq:doireactratefunc}. Thus, instead of the first rate in \cref{eq:diffinflCDMErates}, we use $\lambda_r=\lambda_0 \mathds{1}_{r\leq\sigma}$. In this case the diffusion-influenced reaction rate was calculated in \cite{erban2009stochastic} and is given by
\begin{align}
	k_\text{D}(\lambda_0) = k_S\left[1-\sqrt{\frac{D}{\lambda_0 \sigma^2}}\tanh \left(\sqrt{\frac{\lambda_0 \sigma^2}{D}}\right) \right],
\end{align}
which again in the limit $\lambda_0 \rightarrow \infty$ tends to the diffusion-controlled rate $k_S$. In the formulation of the problem, $C$ is not relevant and $A$ is centered at the origin, so the same result would hold for the system $A+A\rightarrow \emptyset$. Then the question arises: what is the connection between this theory and the formulation of mutual annihilation from \cref{subsec:mutannih}? This question was asked by Doi in \cite{doi1976stochastic}, and it remains a relevant issue. In short, diffusion-influenced reaction theory managed to compress many probabilistic arguments into a simple model, and thus a clear interpretation at the particle level was not trivial \cite{doi1976stochastic,szabo1980first}. 

One can reinterpret the theory in terms of the first reaction time between a pair of particles \cite{szabo1980first}, one fixed at the origin ($A$) and one diffusing ($B$). The inner boundary of the domain ($r=\sigma$) is reactive while the outer boundary ($r=R$) is reflective. One can then calculate the mean first passage time (average time at which $B$ reacts). The rate \cref{eqSmolsrates} is simply the inverse of this mean first passage time as $R\rightarrow \infty$, which is conveniently independent of the initial position. This problem can also be rewritten as a CDME for one diffusive particle with one reactive and one reflective boundary, which is indeed the starting point in \cite{szabo1980first}. As there are no particle interactions except for those emulated by the reactive boundary, the system is linear, so studying the ensemble average of many of these systems is equivalent to studying the mean behavior of the corresponding many-particle system. The probabilistic model based on the CDME from \cref{difflimiCDME} corresponds to this many-particle system, and its mean behavior as $R\rightarrow \infty$ matches that of the first passage time theory \cite{szabo1980first}. However, the models are only equivalent in the limit $R\rightarrow \infty$, which corresponds to the large copy number limit since the number of particles must go to infinity to cover an infinite volume. For finite $R$, the many-particle formulation with a constant concentration at $R$ (\cref{difflimiCDME}) is essentially different to the pair formulation with a reflective boundary condition \cite{szabo1980first}. In our work, we are mainly  interested in the probabilistic dynamics and simulations of open finite-size many particle systems that are coupled to a reservoir, so \cref{difflimiCDME} is the adequate model since it truly models the open many particle system probabilistically. 

Now that we clarified the particle interpretation, one should be able to recover, at least approximately, the diffusion-influenced reaction rate (\cref{eqSmolsrates}) from the equation for the dynamics of the mean concentration of the mutual annihilation example (\cref{eq:RDPDEmutannihcov}) by integrating the spatial degrees of freedom, i.e. 
\begin{align}
\int_\mathbb{X} \left(D\nabla^2 c(y,t) -\lambda_0 \int_{|y-z|\leq \sigma}\mathbb{E}[C(y)C(z)] dz \right) dy \approx - k_\text{D}(\lambda_0)  c(t)^2
\label{eq:RDPDEmutannihcov2}
\end{align}
with $c(t)=\int_\mathbb{X}c(y,t)dy$ as $R\rightarrow \infty$. The resulting equation after integration would be a simple rate equation for the concentration, $\partial_t c(t)=- k_\text{D}(\lambda_0)c(t)^2$ with $k_\text{D}(\lambda_0)$ the effective macroscopic rate. To the best of our knowledge, \cref{eq:RDPDEmutannihcov2} has not been shown analytically, but its convergence can be tested numerically as $R\rightarrow \infty$. For the special case $\lambda_0 \rightarrow \infty$ and assuming an homogeneous system, this was shown by Doi \cite{doi1976stochastic}.

\section{Numerical schemes for reservoir interaction} 
\label{sec:resinteraction}

In this section, we develop numerical schemes to model interactions with a macroscopic reservoir based on the CDME and on the results from \cref{sec:openbnd,sec:BridgingCDME}. As it will be useful for this section, in \cref{sec:reactschemes}, we overview numerical schemes to model reactions with space dependent rates, including zeroth, first and second order reactions as well as absorbing and partially absorbing boundaries. 

The schemes presented in this section are all new developments with exception of the exact explicit scheme from \cref{sec:explicitScheme}, which was first devised in \cite{del2018grand} and improved in \cite{kostre2021coupling}. The beauty of representing the reservoir interaction in terms of ``reaction rate functions'' ---as shown in \cref{sec:openbnd}--- is that we can design novel schemes for the interaction building upon well-known algorithms like the Gillespie algorithm and $\tau$-leaping. Moreover, in \cref{sec:BridgingCDME}, we have shown this representation converges to well-known macroscopic behavior, which guarantees consistency across scales.

Following \cref{eq:CDMEreservoir}, to simulate the interaction with the reservoir through a contact boundary, we need to simulate three processes: the particles leaving the domain into the reservoir, the particles being ``injected'' from the reservoir into the domain and the second (or higher ) order reactions occurring across the contact boundary. The last one we will handle in \cref{sec:reactintfce}, so we first focus in the first two. We already showed the outgoing particles can be modeled with a degradation reaction. However, it is computationally more efficient to simply remove the particle if it diffuses into the reservoir (\cref{alg:absorbingBound}). For the injection of particles, we incorporate a creation reaction with a rate $\lambda_\text{in}$ from \cref{eq:lin_cont}. Note this rate, as well as its discrete counterpart from \cref{eq:lin_discrete,eq:generalRates}, has units of reaction per second per unit volume. As we want the effective rate of incoming particles coming from the mirror boundary layer $\hat{\Omega}$ (\cref{fig:reservoir_interaction}), we multiply \cref{eq:lin_discrete} by its volume $V_{\hat{\Omega}}$,

\begin{equation}
\tilde{\lambda}_\text{in}^\text{eff} = V_{\hat{\Omega}} \tilde{\lambda}_\text{in}(x) =  \gamma_\text{in} n_R,
\label{eq:discreteNumRate}
\end{equation}
where $n_R=V_{\hat{\Omega}} c_R$ is the average number of reservoir particles in the mirrored boundary layer and $\gamma_\text{in}=D/ \Delta x^2$ is the individual particle jumps due to the discretized diffusion. Note $n_R$ does not need to be an integer. 

Computationally, the rate from \cref{eq:discreteNumRate} is accurate for infinitesimal small timesteps. However, it becomes less accurate for larger time steps. To better understand this issue, let us consider the system in a discrete setting. For a given time interval, diffusion limits the number of particles allowed to jump from the reservoir into the system. The maximum number of particles that can jump in one time step from the reservoir into the boundary layer $\Delta x$ is $n_R$ (note the time step should be limited by $\Delta t \leq \Delta x^2/(2D)$ \cite{del2016discrete}). At a given point in time $t=0$, we have the $n_R$ particles in the mirrored boundary layer $\hat{\Omega}$ of the reservoir, each of which can jump with a rate $\gamma_\text{in}$ into the the boundary layer $\Omega$. We want to know: what is the number of incoming/injected particles from the reservoir into the domain up until time $\tau$. We can reformulate this consistently with the CDME as events counted by a Poisson process and write it using the random time-change Poisson representation (see \cref{eq:zerothRTC} and \cite{anderson2015stochastic}),
\begin{align}
N(\tau)= N(0) + Y\left(\int_0^\tau \Lambda(s)ds\right),
\end{align}
where $N(\tau)$ represents the total number of injected particles from the reservoir into the domain during a time interval $\tau$. The  symbol $Y$ represents sampling from a unit rate Poisson process with a rescaled time as argument; in other words $Y$ yields a random number of events---particle injections --- that happen in time interval $\tau$ with rate $\Lambda(t)$. 
At the start of the timestep $t=0$, there are no injected particles, so $N(0)=0$. The first particle injection (first event) has a high probability because any of the $n_R$ particles can jump. This event happens with rate $\gamma_\text{in} n_R$, as in \cref{eq:discreteNumRate}. However, the probability of the second event diminishes because there are only $n_R-1$ particles left, so the corresponding rate is $\gamma_\text{in}(n_R - 1)$, and so on, thus $\Lambda(s) = \gamma_\text{in}(n_R - N(s))$ and
\begin{align}
N(\tau)= Y\left(\int_0^\tau \gamma_\text{in}(n_R - N(s))ds\right).
\label{eq:reservoirRTCP}
\end{align}
Note once $N$ reaches the value $n_R$, the rate remains zero until the end of the timestep thus the condition $N(\tau) \leq n_R$ is always fulfilled. In the next time step, diffusion enables the reservoir to replenish itself and there are again $n_R$ available particles. As we framed our problem using \cref{eq:reservoirRTCP} and consistently with the CDME (\cref{eq:CDMEreservoir}), we can now use a repertoire of methods to solve the interaction with the reservoir. In what follows, we present a few schemes, and we implement them and compare them in \cref{sec:simulations}. 
Extensions to two and three dimensional domains are possible by discretizing the boundary layer in the reservoir  (\cref{fig:diagResInt}) and applying this procedure on each cell (also see \cite{kostre2021coupling}).

\begin{figure}
	\begin{minipage}{0.49\linewidth}
		\begin{algorithm}[H]
			\DontPrintSemicolon
			\SetKwProg{Fn}{Function}{:}{}
			\SetKwFunction{FInjection}{InjectionTauLeap}
			\Fn{\FInjection{$\tau$, M, $\gamma_\text{in}$, $n_R$}}{
				$N \gets 0$ \\
				$\Delta \tau \gets \tau/M$\\ 
				\For{$k=1$ to $M$}{
					$\lambda \gets \gamma_\text{in}(n_R - N)$ \\
					$N \gets N$ + Poisson($\lambda \Delta \tau$)
				}
				AddParticlesFromReservoir(N)
			}
			\vspace{5mm}
			\SetKwFunction{FInjectionG}{InjectionGillespie}
			\Fn{\FInjectionG{$\tau$, $\gamma_\text{in}$, $n_R$}}{
				$N \gets 0$ \\
				$t \gets 0$\\ 
				\While{$t <$ $\tau$ and $N<n_R$}{			
					$r \gets random(0,1)$ \\
					$\lambda \gets \gamma_\text{in}(n_R - N)$\\
					$t \gets t + log(1.0 / r) / \lambda$ \\
					\If{$t \leq \tau$} {
						$N \gets N + 1$ \\
					}
				}
				AddParticlesFromReservoir(N)
			}
			\vspace{2mm}
		\end{algorithm}
	\end{minipage}
	\begin{minipage}{0.49\linewidth}
		\begin{algorithm}[H]
			\DontPrintSemicolon
			\SetKwProg{Fn}{Function}{:}{}
			\SetKwFunction{FInjectionE}{InjectionExplicit}
			\Fn{\FInjectionE{$\tau$, $\gamma_\text{in}$, $n_R$}}{
				$N \gets 0$ \\
				$[n_R] \gets$ floor($n_R$)\\
				$\epsilon \gets n_R - [n_R]$\\
				\textit{react-prob1} $\gets 1-\mathrm{exp}(-\gamma_\text{in} \tau) $ \\
				\textit{react-prob2} $\gets 1-\mathrm{exp}(-\epsilon \ \gamma_\text{in} \tau)$ \\
				\For{$j=1$ to $[n_R] + 1$}{
					$r \gets$ random($0,1$) \\
					\If{$j\leq [n_R]$ and $r<=$ react-prob1}{
						$N \gets N+1$ 
					}
					\ElseIf{$r<=$ react-prob2}{
						$N \gets N+1$
					}
				}
				AddParticlesFromReservoir(N)
			}
			\vspace{5mm}
			\SetKwFunction{FAddParts}{AddParticlesFromReservoir}
			\Fn{\FAddParts{$N$}}{
				\For{$j=1$ to $N$}{
					x $\gets$ Uniform sample in $\Omega$ \\
					add particle at x
				}
			}
		\end{algorithm}
	\end{minipage}
	\caption{Pseudo-code implementations of the reservoir injection procedures from \cref{sec:resinteraction} for one time step $\tau$ in a one-dimensional domain. The first routine corresponds to the injection of particles using the $\tau$-leap algorithm (\cref{eq:tauleap}) with $\tau$ divided in $M$ equal sub-steps. The second and third routine handle the injection with the Gillespie and the explicit exact algorithms. The last routine is auxiliary to the previous three and adds $N$ particles within the boundary layer $\Omega$ next to the reservoir interface. Extensions to two and three dimensions, as well as to time and space dependent reservoirs, are analogous but require discretizing the boundary layer in the reservoir and updating the rate functions accordingly.
	\label{alg:injection}
	} 
\end{figure}

\subsection{Approximate $\tau$-leap scheme:}
\label{sec:tauleapresint}

We can approximate \cref{eq:reservoirRTCP} with the $\tau$-leap algorithm by equating $\tau$ to one time step from the simulation, that is
\begin{align}
N(\tau)= Y\left(\gamma_\text{in}(n_R - N(0))\tau\right) =
Y\left(\gamma_\text{in} n_R \tau \right),
\end{align}
which corresponds to using the discrete rate from \cref{eq:discreteNumRate}. However, we can separate it in $M$ intermediate timesteps to obtain a more accurate numerical scheme, yielding
\begin{align}
N(\tau)= Y\left(\sum_{k=0}^{M-1}\Lambda(\tau_k)(\tau_{k+1}-\tau_k)\right), \qquad \mathrm{with:} \qquad
\Lambda(\tau_k) = 
\begin{cases}
	\gamma_\text{in}(n_R - N(\tau_k)) \quad &\mathrm{if:} \quad N(\tau_k)\leq n_R \\
	0 \quad &\mathrm{otherwise},
\end{cases}
\label{eq:tauleap}
\end{align}
and $0=\tau_0<\tau_1<\cdots<\tau_N = \tau$. A pseudo-code implementation of the algorithm is shown in the \textit{InjectionTauLeap} function in \cref{alg:injection}, and it is implemented in sample cases in \cref{sec:simulations}.

\subsection{Gillespie exact scheme:}
\label{sec:gillespexact}

Alternatively, one can solve \cref{eq:reservoirRTCP} exactly but at a higher computational cost. We can do this by directly applying the Gillespie algorithm to \cref{eq:reservoirRTCP} at each timestep $\tau$. This is done by calculating the time at which the first particle is injected into the reservoir (first reaction time) $t_0$ using the rate $\lambda_0=\gamma_\text{in} n_R$, then calculating the next one $t_1$ using the rate $\lambda_1=\gamma_\text{in} (n_R-1)$ and so on until $\sum t_i$ reaches $\tau$ or the number of injected particles reaches $n_R$:
\begin{itemize}
\item $t_0 = \log(1.0/r_0)/\lambda_0$ with $\lambda_0=\gamma_\text{in} n_R$\\
...
\item $t_i = \log(1.0/r_i)/\lambda_i$ with $\lambda_i=\gamma_\text{in} (n_R - i)$ \\
...
\item Continue until $\displaystyle \sum_{i=0}^N t_i \leq \tau$ or $N=n_R$
\end{itemize}
As the time between events in a Poission process is exponentially distributed, we sample exponential waiting times, so we simply sample a random number $r_i$ between $0$ and $1$ and calculate the exponential waiting time using inverse transform sampling: $t_i = \log(1.0/r_i)/\lambda_i$. The pseudo-code implementation is shown in the \textit{InjectionGillespie} function in \cref{alg:injection}. For a general implementation of the Gillespie algorithm, the reader is referred to \cite{anderson2015stochastic, gillespie1977exact}.

\subsection{Explicit exact scheme:}
\label{sec:explicitScheme}
For completeness, we also present an alternative exact approach that was developed in \cite{del2018grand,kostre2021coupling}. We refer to this as the explicit injection algorithm since we calculate the probability of each particle to jump explicitly. We first separate the average number of particles $n_r$ into its integer and decimal part $n_R=[n_R] + \epsilon$, where $[\cdot]$ denotes the floor function and $\epsilon$ the decimal part. Then, for a timestep $\tau$, the probability for each one of the $[n_R]$ particles to jump with rate $\gamma_\text{in}$ is analogous to \cref{eq:probReact}
\begin{align}
p_1( \gamma_\text{in}, \tau) &= 1 - e^{-\gamma_\text{in} \tau}. 
\end{align}
Then we sample $[n_R]$ random numbers uniformly on $(0,1)$. From these, only $N_1$ of them will be smaller than $p_1$, which corresponds to $N_1$ particles being injected to the system. To handle the decimal contribution, we first need to calculate the probability of one more particle jumping by scaling the rate by $\epsilon$, i.e.
\begin{align}
p_\epsilon(\epsilon \ \gamma_\text{in} , \tau) &= 1 - e^{-\epsilon \ \gamma_\text{in} \tau}.
\end{align}
Then we sample one more random number uniformly on $(0,1)$ and check if it is less than $p_\epsilon$. If so $N_\epsilon=1$, otherwise $0$, and thus the total number of particles jumping from the reservoir into the system in a timestep $\tau$ is  
\begin{align}
N(\tau) = N_1 + N_\epsilon.
\end{align}
This algorithm is also exact as the Gillespie algorithm applied to \cref{eq:reservoirRTCP}, and it is consistent with the methodologies presented in \cite{del2018grand,kostre2021coupling}. However, it requires sampling around $[n_R]$ random numbers at each timestep, while the $\tau$-leap implementation often works well with a few sub-steps (i.e. $5$ to $10$ sub-steps require sampling $5$ to $10$ random numbers). The pseudo-code implementation is shown in the \textit{InjectionExplicit} function in \cref{alg:injection}, and the comparison between the algorithms is shown in \cref{sec:simulations}.

\begin{figure}[bt]
	\centering
	\textbf{a.} \includegraphics[width=0.4\textwidth]{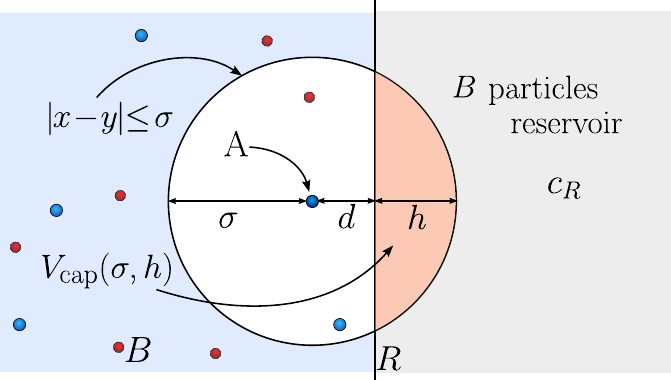}
	\quad \textbf{b.} \includegraphics[width=0.4\textwidth]{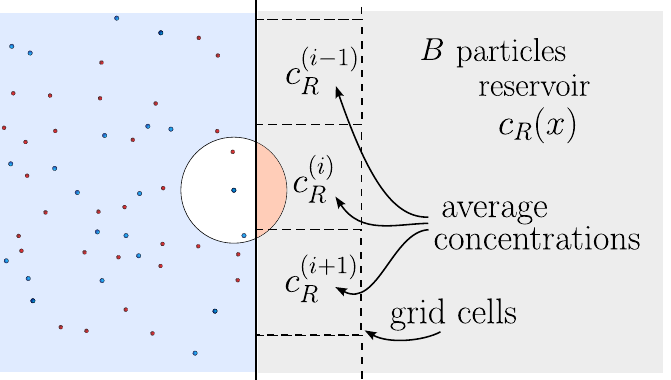}
	\caption{Diagrams of numerical schemes to handle reactions at the reservoir interface. The system domain is illustrated in blue and the reservoir region in grey. \textbf{a.} Illustration of the intersection of the volume reactivity region with the reservoir for the reaction $A+B\rightarrow \emptyset$ used in \cref{eq:annihRateBimoc}. \textbf{b.} Illustration of one possible scheme to handle reservoirs with space-dependent concentrations. It requires setting grid cells along the contact boundary and calculating the average concentration on each grid cell ($c_R^{(i)}$ is the average for cell $i$). In practice, the grid cells are considerably larger than the diameter $2\sigma$ (see \cref{sec:simulationsB}). Note this same discretization can be used for the two (and three dimensional) implementations of the injection procedures.}
	\label{fig:diagResInt}
\end{figure} 

\subsection{Reactions at the reservoir interface (implicit method)}
\label{sec:reactintfce}

We have presented schemes to handle coupling of the diffusion process with the reservoir, but we haven't specified how to handle reactions occurring across the contact boundary. Here, we develop numerical schemes for this setting following \cref{sec:reactInterface}. As zeroth and first-order reactions depend on zero or one particle, we do not require any additional consideration at the interface. Thus, we focus on second-order reactions, where a reactant in the particle domain can react with the other reactant in the reservoir.

We follow the setup from \cref{sec:reactInterface} for the reaction $A +B\rightarrow \emptyset$ with rate function $\lambda(x,y)$ and a reservoir of $B$ particles with fixed concentration $c_R$. The three-dimensional frame of reference is given by the coordinates $x=(x_1,x_2,x_3)$, or by $(x_1,x_2)$ or simply $(x_1)$ if it is only two or one-dimensional. We assume the reservoir is in contact with the system on the right boundary at $x_1=R$. According to \cref{eq:CDMEresreact}, we can model the particles in the reservoir implicitly, and we only need to incorporate a degradation reaction $A\rightarrow \emptyset$ with rate function $\hat{\lambda}(x)$ given by \cref{eq:ratefuncAtInterface}. As often the case in simulations, we assume the volume reactivity model for bimolecular reactions from \cref{eq:DoiRate} such that $\lambda(x,y)=\lambda_0 \mathds{1}_{|x-y|\leq \sigma}$, then \cref{eq:ratefuncAtInterface} yields
\begin{align}
	\hat{\lambda}(x) &=c_R \lambda_0\int_\mathcal{R} \mathds{1}_{|x-y|\leq \sigma} dy, \notag\\
	&=c_R \lambda_0 V_\text{cap}(\sigma,h),
	\label{eq:annihRateBimoc}
\end{align}	
where $V_\text{cap}(\sigma,h)$ is the volume of a spherical cap with height $h$ and radius $\sigma$, $h=\sigma-d$ with $d$ the shortest distance between the particle and the reservoir interface, see \cref{fig:diagResInt}a. In three dimensions, $V_\text{cap}(\sigma,h)=\pi h^2 (3\sigma - h)/3$, in two dimensions $V_\text{cap}(\sigma,h)=\sigma^2\arccos [(\sigma-h)/\sigma] - (\sigma-h)\sqrt{2\sigma h-h^2}$ and in one dimension $V_\text{cap}(\sigma,h)=h$. Thus, to model the bimolecular reaction at the boundary, we simply annihilate every $A$ particle with the rate from \cref{eq:annihRateBimoc}, whenever it is a distance $d\leq \sigma$ from the contact boundary with the reservoir. If there are other $B$ particles within the domain and inside the reactive region, we can choose randomly which reaction occurs or use an event manager \cite{hoffmann2019readdy}. 

There are more complex scenarios to take into account. For reactions with products, the products are only incorporated if the product sampled location is within the system domain. For more complex boundary geometries, one has to consider \cref{eq:CDMEresreact,eq:ratefuncAtInterface} as a starting point. Moreover, if the reservoir concentration is also dependent on time one has to update the rate at every time step. Finally, if the reservoir concentration is space dependent, instead of \cref{eq:ratefuncAtInterface}, we need to approximate the integral 
\begin{align}
	\hat{\lambda}(x) = \int_\mathcal{R}c_R(y) \lambda(x,y) dy.
	\label{eq:ratefuncAtInterface2}
\end{align}	
To avoid calculating this quantity for every $A$ particle, a simple and effective approach is to discretize the region along the contact boundary on the reservoir domain in vertical grid cells, then calculate the average concentration of the reservoir in each of these cells and use the concentration of the grid cell closest to the reacting particle as an approximation of $c_R(y)$ in the integral, see \cref{fig:diagResInt}b. Then, we proceed with the method as before. In general, the size of the grid cells is chosen to be considerably larger than the radius $\sigma$, so this yields a good approximation. Alternatively, one can develop more sophisticated approximations of the integral and reaction rate functions in \cref{eq:ratefuncAtInterface2} resulting in new and more accurate methods.  Note this approach, although often not exact, is considerably more efficient and simpler to implement than the one presented originally in \cite{kostre2021coupling}, which requires explicitly adding virtual particles in the reservoir domain. Thus, we refer to the method just presented as the implicit method to handle reactions at the reservoir interface, where else the method presented in \cite{kostre2021coupling} corresponds to the explicit method.

\section{Simulations}
\label{sec:simulations}

To showcase the numerical methods from \cref{sec:resinteraction}, we apply them to two examples. The first one revisits the topic of diffusion-influenced reactions from \cref{sec:diffinflCDMElimit}, but this time we do a numerical particle-based implementation, which also extends and generalizes the schemes developed in \cite{del2018grand}. The second one, consists of a hybrid simulation between a particle-based system and a PDE-mediated reservoir for an SIR epidemics model, resulting in new and more efficient versions of the methods presented in \cite{kostre2021coupling}.

\subsection{Open diffusion system with reactive boundary and in contact with a reservoir}
\label{sec:simulationsA}

\begin{figure}[tb]
	\centering
	\textbf{a.} 
	\includegraphics[width=0.23\textwidth]{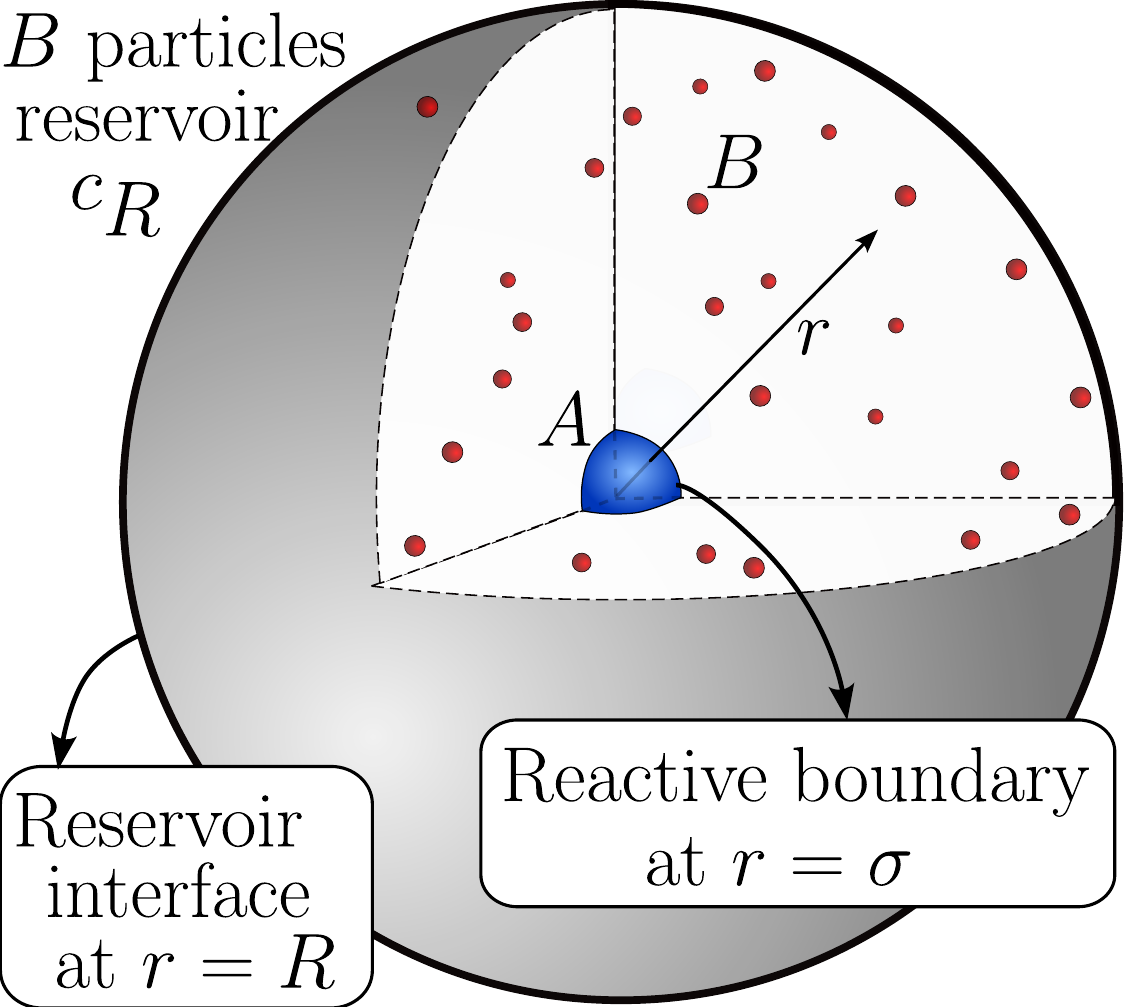}
	\textbf{b.} 
	\includegraphics[width=0.66\textwidth]{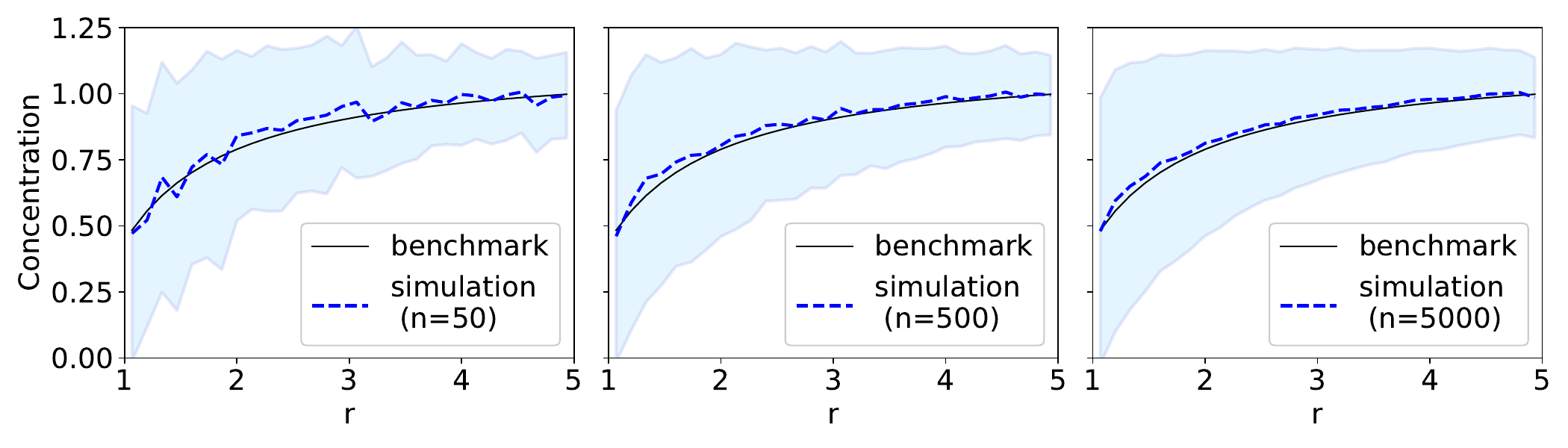}\\
	\textbf{c.}
	\includegraphics[width=0.27\textwidth]{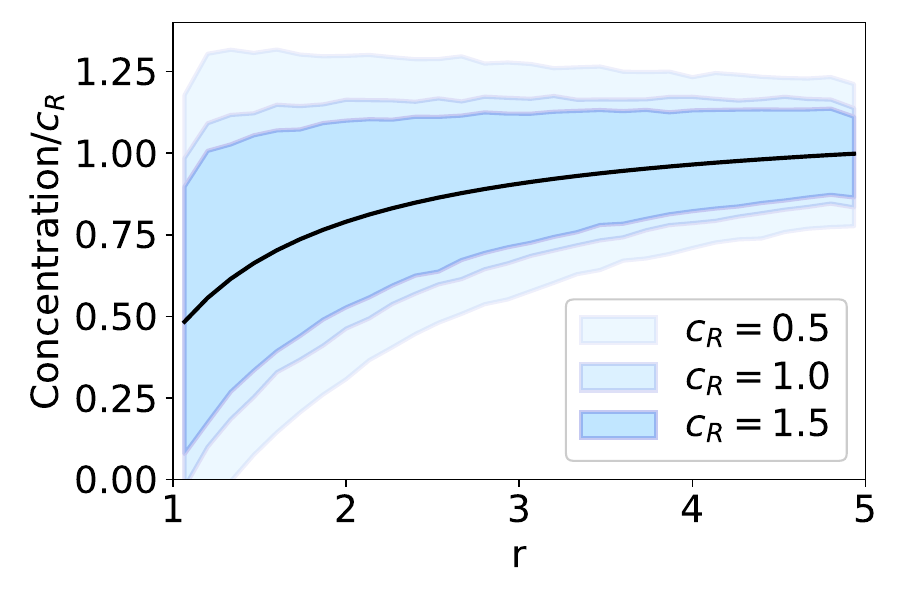}
	\textbf{d.}\includegraphics[width=0.3\textwidth]{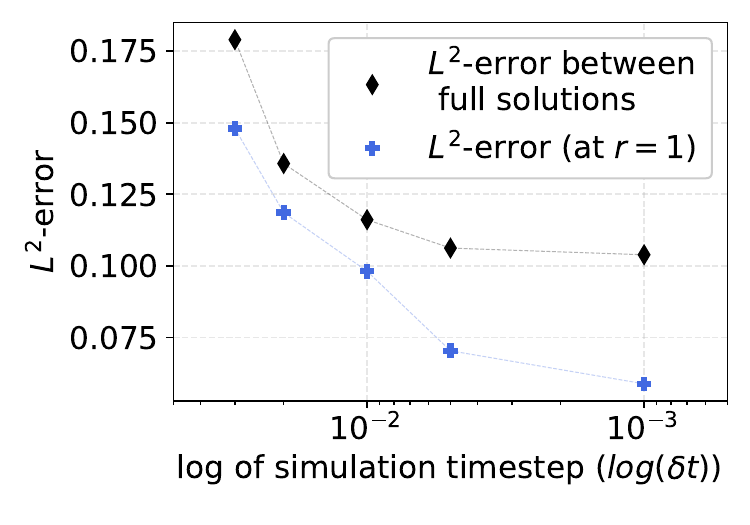} 
	\textbf{e.}\includegraphics[width=0.3\textwidth]{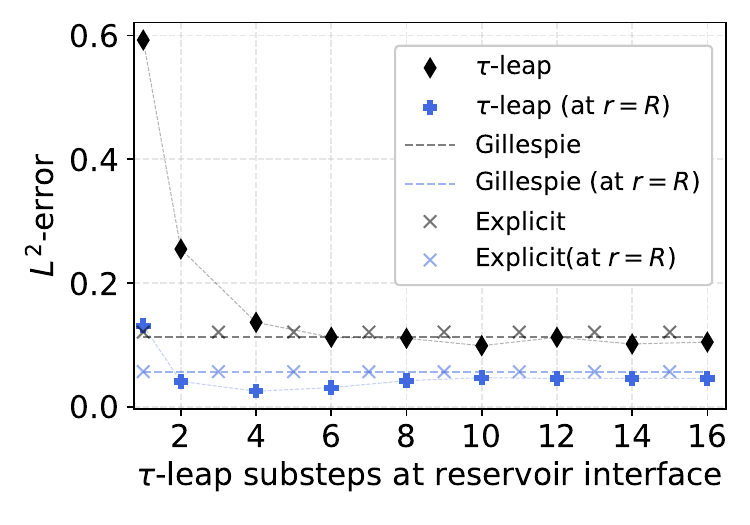}
	\caption{
		Simulation and comparison of the open diffusion system from \cref{sec:simulationsA}. \textbf{a.} Diagram of the particle-based simulation setup. \textbf{b.}
		Comparison between the concentration profile of many average particle-based simulations and the mean-field (solution of \cref{eq:smoleq,eq:smoleqBC}) as a function of the radial distance $r$. The averages were calculated over $n=100,500$ and $5000$ simulations, showing convergence as $n\rightarrow \infty$. The shaded region is the standard deviation. The simulation parameters are: $D=0.5$ (diffusion coefficient), $\sigma=1$, $R=5$, $\kappa=10$, $c_R=1.0$ and time step $\Delta t = 0.001$. The reservoir interaction uses the $\tau$-leap approximate scheme with $10$ substeps. \textbf{c.} Comparison of relative standard deviation for three different reservoir concentrations. The concentrations were averaged over $5000$ simulations and normalized over $c_R$. As $c_R$ grows, the number of particles in the system grows and the standard deviation is reduced, showing convergence through the law of large numbers.
		\textbf{d.} Semi-log plot of $L^2-$norm errors between the steady state analytic solution of the PDE (\cref{eq:smoleq}) and the average concentration from $1000$ particle-based simulations as a function of $\Delta t=0.03,0.02,0.01,0.005,0.001$. The parameters are the same as in b., except for a smaller domain with $R=1$ and the $\tau$-leap substeps is fixed to $10$. The blue crosses show the same error but only using the two values closest to $r=1$. \textbf{e.} Same error as in d. using a fixed timestep of $\Delta t=0.01$ and as a function of the number of $\tau-$leap substeps. The blue crosses represent the same error at the two values closest to $r=R$. As reference, the errors using the Gillespie and the explicit exact algorithms are also shown. Note we expect small error fluctuations and that both errors plateu due to the intrinsic error of averaging over a fixed number of simulations.
	}
	\label{fig:smol_convergence}
\end{figure} 

We follow the same setup as in \cref{sec:diffinflCDMElimit} and construct a simulation that is consistent with \cref{difflimiCDME} (\cref{fig:smol_convergence}a). Assume an arbitrary and variable number of $B$ particles diffuse in a three-dimensional domain delimited by the spherical region with radius $r\in[\sigma,R]$. According to \cref{difflimiCDME}, the \textit{diffusion} follows standard Brownian motion and is implemented with the procedure in \cref{alg:reactions}. To model the reactions, we implement a partially absorbing boundary at $r=\sigma$ with rate $\kappa$ by following the implementation from \cref{sec:PRBM}. In the far-field $r=R$, the system is in contact with a reservoir with constant concentration $c_R$, so one needs to model the \textit{injection} of particles following \cref{sec:resinteraction}. The diffusion, reaction and reservoir interaction can occur simultaneously and possibly at different timescales. To obtain a robust and stable numerical algorithm it is convenient to implement a Strang splitting (\cite{kim2017stochastic,strang1968construction,kostre2021coupling,del2018grand}) at each time step $\tau$ in the following form 
\begin{enumerate}
\setlength\itemsep{-1.0em}
\item Injection($\tau/2$) \ and \
PartiallyAbsorbingBoundary($\tau/2$) \\
\item Diffusion($\tau$) $\rightarrow$ Enforce boundary conditions (reflective and absorbing) \\
\item Injection($\tau/2$) \ and \ PartiallyAbsorbingBoundary($\tau/2$) 
\end{enumerate}
We can use any of the specific routines from \cref{alg:injection,alg:absorbingBound,alg:reactions}.
Note as the injection happens at the far boundary and the 
reactions at the near one, the order in which these routines are applied in steps one and three is not important.
Independently of the partially absorbing routines and injection routines, we need to impose boundary conditions to the diffusion process itself. At $r=\sigma$, the system has a reflective boundary condition since the absorption is uniquely handled by the \textit{PartiallyAbsorbingBoundary} routine. The reflective boundary is enforced by rejection sampling, i.e. only diffusion steps that end with positions $r\geq \sigma$ are allowed. In the boundary in contact with the reservoir at $r=R$, the flow of particles into the reservoir is enforced by eliminating any particle that diffuses beyond $r\geq R$, while the influx of particles is handled by the \textit{Injection} routine.

Due to the spherical symmetry of the domain, we can handle the system as one-dimensional. However, the injection of particles needs to take into account the spherical boundary.
In \cite{del2016discrete,del2018grand}, it is shown how the diffusion discrete rates are corrected for a spherical boundary. In this case, one possible simple discretization is $\gamma_\text{in} = D/\Delta r^2 - D/(\Delta r R)$ in \cref{eq:generalRates}. If one wants to model the outflux as a reaction too, one would use the value $\gamma_\text{out} = D/\Delta r^2 + D/(\Delta r R)$ in \cref{eq:generalRates}. Similar approaches can be used for more complex geometries \cite{peskin2002immersed,li2006immersed}.

\Cref{fig:smol_convergence}b shows the results of the particle-based simulations using this setup. The reservoir interaction is modeled using the $\tau$-leap version of the scheme to improve efficiency. The concentration is shown as function of the radial distance $r$ averaged over different numbers of simulations and compared with the analytic solution of \cref{eq:smoleq,eq:smoleqBC}. 

To understand how to choose the time step $\Delta t$ and the number of $\tau$-leap substeps in the simulations, we investigated the $L^2$-error in terms of these parameters. We calculate this error for both the solution in the whole domain and just along the boundary.
\Cref{fig:smol_convergence}d shows the $L^2$-error between the average over $1000$ particle-based simulations and the mean-field PDE solution as a function of the simulation time step $\Delta t$. \Cref{fig:smol_convergence}e shows the same $L^2$-error for different number of $\tau$-leap substeps for the injection routine. The errors produced by the Gillespie and the explicit exact algorithms from \cref{sec:resinteraction} are also shown for reference. From these figures, we observe a time step of $\Delta t =0.01$ is quite good but $\Delta t=0.001$ is even better if we want good accuracy at the boundary $r=\sigma$. We also observe that for the $\tau$-leap scheme to reach a similar accuracy to that of exact algorithms, anything above $6$ substeps is already quite good for this example. With this information, we chose the parameters for the simulations presented in \cref{fig:smol_convergence}b, $\Delta t=0.001$ and $10$ $\tau$-leap substeps. The rest of the parameters are shown in the picture caption.

The results from \cref{fig:smol_convergence} show remarkable consistency between the mean behavior of particle-based simulations and the expected mean-field behavior of the open system. Similar simulations were presented in \cite{del2018grand} using solely an old version of the explicit scheme for the reservoir interaction. Here, by capturing the reservoir interaction into rates within the CDME (\cref{sec:openbnd,sec:diffinflCDMElimit}), we have devised a robust analytical framework to construct consistent numerical schemes for open reaction-diffusion systems. With this framework, we not only recover the previous schemes, but we also enable the development of new schemes for reservoir interaction with improved efficiency, such as the $\tau$-leap implementation of the injection procedure (\cref{sec:tauleapresint}). However, note that for small reservoir concentrations (e.g. corresponding to an average injection of around 5 or less particles per timestep), the exact injection schemes (\cref{sec:gillespexact,sec:explicitScheme}) will remain more accurate and perhaps even more efficient than the $\tau-$leap implementation (e.g. with $5$ substeps). 

We presented the coupling with a reservoir with constant concentration and with no reactions occurring across the reservoir interface, and we also compared different injection algorithms. In the next example, we will cover a more complex implementation.

\subsection{Hybrid SIR simulation}
\label{sec:simulationsB}
\begin{figure}
	\centering
	\includegraphics[width=0.9\textwidth]{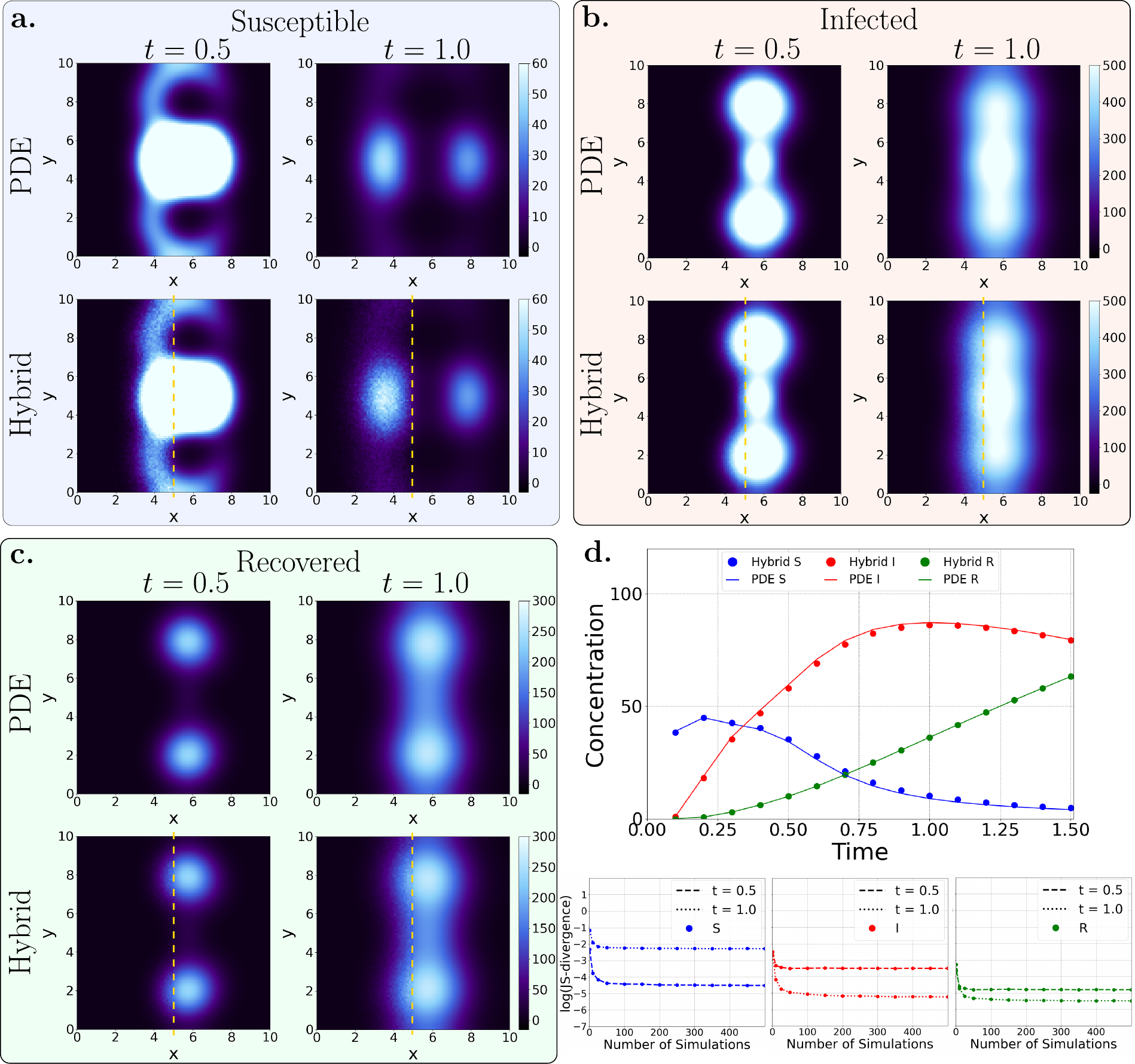}
	\caption{Hybrid simulations comparisons for the SIR model using  and the scheme from \cref{sec:reactintfce}. The parameters are $D_S, D_I, D_R=0.8$, $\kappa=0.015$, $\beta=0.5$, timestep $\Delta t=0.005$ and initial conditions as described in the text. The reaction radius for the particle simulations is $\sigma=0.01$ \textbf{a.} Comparison of the concentration of ``suceptible'' agents at two times. For reference, we show the numerical solution of the PDE in the whole domain. The hybrid simulation uses the PDE solution on the right half of the domain ($[5,10) \times [0,10]$), while the left side ($[0,5) \times [0,10]$) results from an average over $500$ particle-based simulations. The division between the two domains is shown with a yellow dashed line. \textbf{b.} and \textbf{c.} Same comparison for ``infected'' and ``recovered'' agents. \textbf{d.} Comparison of the space averaged concentration in the particle domain as a function of time, lines represent the PDE reference solution, dots represent the solution using the hybrid scheme. Below, the Jensen-Shannon (JS) divergence for susceptible, infected and recovered agents is shown in a logarithmic scale at two times. The x-axis is the number of particle-based simulations used to calculate the average. Each point is calculated using 50 bootstrapped samples.}
	\label{fig:reaction_solutions}
\end{figure}

In this section, we extend the previous implementations to work with reservoirs that have time and space-dependent concentrations. We model the reservoir dynamics with a reaction-diffusion PDE, which is then coupled to the particle-based simulation, resulting on a hybrid scheme. An earlier version of this scheme was presented in \cite{kostre2021coupling}. However, this version has a significant simpler implementation emerging from the theoretical results of this paper, and it is significantly more efficient. To exemplify the scheme implementation, we consider the SIR model for spread of disease given by two ``chemical reactions'', infection and recovery:
\[
\mathcal{S}+\mathcal{I}\rightarrow 2\mathcal{I}, \qquad
\mathcal{I}\rightarrow \mathcal{R}
\]
with susceptible $\mathcal{S}$, infected $\mathcal{I}$, and recovered agents $\mathcal{R}$. If these agents diffuse in space following standard Brownian motion, the corresponding reaction-diffusion PDEs are given by
\begin{align}
	\label{eq:reaction_diffusionSIRExample}
	\partial_t c_s(x,t) & = D_S\nabla^2 c_s(x,t) - \kappa \cdot c_s(x,t) \cdot c_i(x,t) \notag \\
	\partial_t c_i(x,t) & = D_I\nabla^2 c_i(x,t) + \kappa \cdot c_s(x,t) \cdot c_i(x,t) - \beta \cdot c_i(x,t) \\
	\partial_t c_r(x,t) & = D_R\nabla^2 c_r(x,t) + \beta \cdot c_i(x,t), \notag
\end{align}
where $\kappa$ and $\beta$ are the infection and recovery rates, and $D_S$, $D_I$ and $D_R$ the diffusion coefficients respectively. We consider a two-dimensional domain $[0,10] \times [0,10]$, which is divided into two halves at $x=5$. The left part corresponds to the particle domain and  the right one to the reservoir/concentration domain.  We initialized the populations of susceptible and recovered individuals by distributing them across specific regions in the domain. For the initial condition, the concentration of susceptible agents is concentrated within three square areas of length $1\times 1$ centered at $(5.6,2)$, $(5.6,5)$ and $(5.6,7)$ with a uniform concentration of $7000$. Similarly, the infected individuals were placed in two circular regions of radius $0.5$ centered at $(6.5,2.5)$ and $(6.5,7.5)$ with a uniform concentration of $800$. The setup was devised this way to showcase the capability of the scheme, and the accuracy of the coupling at the interface between the particle domain and the reservoir domain.

To incorporate reactions into the coupling scheme, ensuring stability, we applied second-order Strang splitting in the following three steps:
\begin{enumerate}
	\item Injection \(\left(\tau/2\right)\)
	$\rightarrow$ Reactions \(\left(\tau/2\right)\)
	\item Diffusion \(\left(\tau\right)\) $\rightarrow$ Enforce boundary conditions (reflective and absorbing)
	\item Reactions \(\left(\tau/2\right)\) 
	$\rightarrow$ Injection \(\left(\tau/2\right)\)
\end{enumerate}

To model the reactions occurring across the contact boundary between the reservoir and the particle domain, we implement the scheme from \cref{sec:reactintfce}. To focus on the accuracy of this coupling, the injection of particles is modeled with an exact scheme, we chose the one from \cref{sec:explicitScheme}, but we could have also chosen \cref{sec:gillespexact}. To accurately model the infection and recovery processes in the domain and in the boundary, we applied additional Strang splitting in each of the reaction steps, Reactions $\left(\tau/2\right)$, as follows: recovery $\left(\tau/4\right)$, infection $\left(\tau/2\right)$ and recovery $\left(\tau/4\right)$.

Figure \ref{fig:reaction_solutions} compares the concentration profile between the PDE and the hybrid simulation at time-steps $t=0.5$ and $t=1.0$ of the susceptible (a), infected (b), and recovered (c) species. The reference solution corresponds to the reaction-diffusion PDE solved in the whole domain using standard finite difference techniques \cite{kostre2021coupling, leveque2007finite}. The hybrid simulation consists of the right half of the domain solved by the same PDE, while the left half of the domain corresponds to a particle-based simulation coupled to the PDE domain using our hybrid scheme. The concentration shown in this region corresponds to the average concentration over $500$ particle simulations. We observe an excellent agreement between the mean-field behavior of the particle system and the PDE, validating the simulation consistency across scales. 

We further verify the consistency of the hybrid simulation explicitly in \cref{fig:reaction_solutions}d, where we calculate the space-averaged concentration of individuals in the particle domain and compare it with the space-averaged concentration of the PDE. The lines represent the PDE reference solutions; the dots represent the solution using the hybrid scheme. We observe a good match between all solutions. For this simulation, the approximate schemes for the hybrid simulation were eight times more time efficient than the explicit exact ones developed in our previous work \cite{kostre2021coupling}.   Finally, as a last validation test, we calculated the Jensen-Shannon (JS) divergence between the reference solutions and the hybrid solutions (for two time steps) averaged over different numbers of simulations, as shown in the three bottom plots of \cref{fig:reaction_solutions}d. The JS-divergence decreases monotonically with increasing number of simulations illustrating the convergence towards the mean-field described by the PDE model. The code and software developed for this work is open source and available under an MIT license in \href{https://github.com/MargKos/multiscaleRD}{github.com/MargKos/multiscaleRD} and Zenodo \cite{kostreMultiscaleRD}.

\section{Concluding remarks}
To summarize, we extended the CDME ---the master equation for the probabilistic dynamics of reaction-diffusion processes at the particle level--- to incorporate interactions with macroscopic material reservoirs. This is achieved by emulating the flux of particles into and out of the reservoir through creation and degradation reactions using localized rate functions with a specific form. We further presented a methodology to recover macroscopic concentration-based descriptions from particle-based probabilistic descriptions. This consists of taking an expectation on the CDME that yields an equation for the average concentration field. We showcased the methodology for linear and nonlinear reactions, as well as for an archetypal model of an open reaction-diffusion system: diffusion-influenced reactions. The linear case is relatively straightforward, while the nonlinear case requires additional considerations to obtain closed equations, such as restricting the form of the rate function and taking large copy number limits. Both recovered well-known reaction-diffusion PDEs for the evolution of the concentration, as previously shown with alternate methods. For the diffusion-influenced reactions case, we built the particle-based description based on the CDME with reservoir interaction developed at the start of this work. The resulting concentration-based model matches the well-known macroscopic model, Smoluchowski's equation, with its corresponding boundary conditions. This explicit connection between the underlying CDME at the particle level and the Smoluchowski's model, by using the ``reaction rates'' as the coupling mechanism with the reservoir, has not been presented before. We further used these results to discussed how the CDME can be used to interpret other formulations of diffusion-influenced reactions. Finally, to show the practical utility of the theoretical results, as well as their numerical validation, we derived and implemented numerical schemes for two relevant examples. This not only provided a theoretical robust justification of previous schemes for similar problems \cite{del2018grand,kostre2021coupling} but also enabled the derivation of novel schemes with increased efficiency and easier implementation, as presented in this manuscript. The results in this work thus bridge probabilistic theory and simulations for open reaction-diffusion systems across scales, laying the foundations for a multiscale framework for reaction-diffusion processes.

The framework presented not only connects microscopic probabilistic models at the particle level with macroscopic deterministic concentration-based models but also yields connections between the parameters at the different scales. The bridging between the micro and the macroscopic scales requires taking certain limits and approximations. By imposing restrictions on these approximations, the framework has the potential to yield mesoscopic models. For instance, in the case of mutual annihilation of \cref{subsec:mutannih} (or for nonlinear reactions in general) one could keep the covariance term in \cref{eq:RDPDEmutannihcov} and try to approximate it. Alternatively, one can perform a moment expansion of the CDME, perhaps up to second order, to account for the covariances. Such a model could be re-framed in the form of a stochastic PDE, which works at the concentration level but preserves information about fluctuations and covariances.

As shown in \cref{sec:resinteraction,sec:simulations}, this work has natural applications in the development of numerical schemes for open complex systems, such as biochemical reaction systems (see also \cite{anderson2015stochastic,del2021multiscale,dibak2018msm,lester2015adaptive,smith2018auxiliary,smith2018spatially, winkelmann2020stochastic}). The CDME provides a ground truth for the particle-based model from which one can derive coarser models, yielding mathematical bridges between the models and parameters at different scales. Based on these ``bridges'', one can construct multiscale simulation schemes for reaction-diffusion processes that are consistent across scales. This is precisely what we did in \cref{sec:resinteraction,sec:simulations}. However, the application scope to numerical schemes goes beyond the ones presented in this manuscript. For instance, one can apply a Galerkin discretization to the CDME \cite{del2021probabilistic} and recover the reaction-diffusion master equation (as similarly done in \cite{isaacson2013convergent}) or the spatiotemporal master equation \cite{winkelmann2016spatiotemporal}, both spatially discretized forms of the CDME. As the discretization is built based on the underlying CDME, one organically obtains a convergent equation in the continuous limit that enables constructing multiscale schemes with adaptive resolution. Another example is on coupling reactions occurring across contact boundaries, i.e. the approximation of the integral in \cref{eq:ratefuncAtInterface2}. Our scheme resulted from specific assumptions and a simple approximation to this integral. However, one can perform more sophisticated approximations yielding novel numerical schemes. One last example consists of using rate functions that depend on orientation or other internal variables. We can then observe how these dependencies manifest in emergent macroscopic models, enabling once again multiscale simulations in novel settings. In all these examples, the connection between numerical schemes and the CDME becomes essential to construct accurate numerical schemes.

As a closing remark, the multiscale framework/workflow introduced here can be extended to complex systems beyond reaction-diffusion. As mentioned in \cite{del2025dynamics}, one can extend the CDME to a master equation for general dynamical systems with varying particle number. For instance, the diffusion part can model classical physical particle systems by incorporating the velocity into the dynamics, while the reaction part can model any process that changes the particle number, such as reactions or reservoir interactions. This would yield a master equation for Langevin dynamics with varying particle number. Although perhaps such formulation is not novel per se, one can then use the results of this work to couple the system to dynamic reservoirs by employing velocity-dependent rate functions with potential applications to molecular dynamics and fluctuating hydrodynamics. Another example is in social agent-based phenomena, which rarely occurs in isolation from the environment. In this case, the CDME can be extended to handle the transport dynamics of agents in real or abstract space (e.g. opinion space \cite{helfmann2023modelling}) and where the number of agents is not fixed due to continuous influx and outflux of agents into the system. In such cases, one can in principle apply similar procedures, yielding multiscale models and simulations that are consistent across scales.

\begin{acknowledgments}
M.J.R. is supported by the Deutsche Forschungsgemeinschaft (DFG) grant no. RA 3601/1-1. M.K. is funded by the Deutsche Forschungsgemeinschaft (DFG, German Research
Foundation) under Germany´s Excellence Strategy – The Berlin Mathematics
Research Center MATH+ (EXC-2046/1, project ID: 390685689). The authors also thanks Stefanie Winkelmann, Felix Höfling and Luigi Delle Site for insightful discussions.
\end{acknowledgments}

\appendix

\section{CDME to concentration descriptions}
\label{sec:CDMEtoconc}

We can recover the mean concentration from the CDME for one species using the following formula (see eq.~(115) in ref.~\onlinecite{del2021probabilistic}):
\begin{align}
c(y) &= \mathbb{E}[C(y)]=  \lim_{\epsilon \rightarrow 0} \frac{1}{\text{vol}(B_\epsilon)(y)}\mathbb{E}[N_B(y)] \notag \\
&= \sum_{n=1}^\infty n \int_{\mathbb{X}^{n-1}} \, \rho_n(y, x^{(n-1)}) \, dx^{(n-1)}.
\label{eq:conc}
\end{align}
This equation is derived in \cite{del2021probabilistic} by calculating the expected number of particles $\mathbb{E}[N_{B}(y)]$ in a ball of radius $\epsilon$ centered at $y$: $B_\epsilon(y)$. Then one divides by the volume of the ball and takes the limit $\epsilon\rightarrow  0$, which is the average molecular concentration at point $y$. The same equation can be obtained by taking expectations of the stochastic number density fields expressed as sum of delta functions, e.g. see \cite{doi1976stochastic} and the proof of proposition 6.6 from \cite{isaacson2022mean}. One can similarly obtain the covariances
\begin{align}
\mathbb{E}[C(y)C(z)]=
& \sum_{n=2}^\infty n(n-1) \int_{\mathbb{X}^{n-2}} \, \rho_n(y,z, x^{(n-2)}) \, dx^{(n-2)}.
\label{eq:conccov}
\end{align}
Note in all calculations below we often omit the explicit dependence of $\rho_n(x^{(n)},t)$ and $c(y,t)$ on time to simplify notation.

\subsection{Degradation}

Let us consider the degradation reaction $A\rightarrow \emptyset$. The corresponding CDME reads
\begin{align}\begin{split} \label{eq:CDME_degradation}
	\partial_t \rho_n (x^{(n)}) &=
	\sum_{i=1}^n D_i \rho_n(x^{(n)}) -\sum_{i=1}^n \lambda_d(x^{(n)}_i)\rho_n(x^{(n)}) + (n+1)\int_{\mathbb{X}} \lambda_d(y)  \rho_{n+1}(x^{(n)},y) \, dy \,.
\end{split}\end{align}

Following \cref{eq:conc}, we can integrate and apply the sum to obtain an evolution equation for the mean concentration. For simplicity, we omit time dependence from the notation in the calculations below, and instead of $(x^{(n)})$, we often write $(y,x^{(n-1)})$. The left-hand side is 
\begin{align}
\sum_{n=1}^\infty n \int_{\mathbb{X}^{n-1}} \partial_t \rho_n(y,x^{(n-1)})dx^{(n-1)} = \partial_t c(y).
\end{align}
The diffusion term simplifies as follows
\begin{align} \begin{split}
	\sum_{n=1}^\infty n &\int_{\mathbb{X}^{n-1}} \mathcal{D}_n \rho_n(y,x^{(n-1)})dx^{(n-1)} \\
	&= \sum_{n=1}^\infty n \int_{\mathbb{X}^{n-1}} \sum_{i=1}^n D_i \rho_n(y,x^{(n-1)})dx^{(n-1)} \\ 
	&= \sum_{n=1}^\infty n \int_{\mathbb{X}^ {n-1}} D_1 \rho_n(y,x^{(n-1)})dx^{(n-1)}  + 
	\sum_{n=1}^\infty n \int_{\mathbb{X}^{n-1}} \sum_{i=2}^n D_i \rho_n(y,x^{(n-1)})dx^{(n-1)}\\
	&=\mathcal{D} c(y) + \sum_{n=1}^\infty n \int_{\mathbb{X}^{n-1}} \sum_{i=2}^n D_i \rho_n(y,x^{(n-1)})dx^{(n-1)},
\end{split}
\end{align} 
where $D_i$ is the diffusion operator acting on the $i$th component of $\rho_n$. Assuming $D_i\rho_n = D \nabla \cdot \nabla_i \rho_n$, we can apply Gauss theorem to the core of the second term
\begin{align}\begin{split}
	\int_{\mathbb{X}^{n-1}}  D \nabla \cdot \nabla_i \rho_n(y,x^{(n-1)})dx^{(n-1)} =
	\int_{\partial \mathbb{X}^{n-1}} D \nabla_i \rho_n(y,x^{(n-1)}) \cdot \hat{\nu} \ dS. 
\end{split}
\end{align}
Assuming there is no flux across the domain boundary, this term is zero and the diffusion is simply $\mathcal{D} c(y,t)=D\nabla^2 c(y,t)$. Applying the same operations to the reaction terms, we obtain
\begin{align}\begin{split}
	\sum_{n=1}^\infty n & \int_{\mathbb{X}^{n-1}}	\left( (n+1)\int_{\mathbb{X}} \lambda_d(z)  \rho_{n+1}(y,z,x^{(n-1)}) \, dz -\sum_{i=1}^n \lambda_d\big((y,x^{(n-1)})_i\big)  \rho_n(y,x^{(n-1)}) \right) dx^{(n-1)} \\
	& =\sum_{n=1}^\infty n(n+1)  \int_{\mathbb{X}^{n}}	 \lambda_d(z) \rho_{n+1}(y,z,x^{(n-1)}) \,  dzdx^{(n-1)} \\
	& \qquad- \sum_{n=1}^\infty n \int_{\mathbb{X}^{n-1}} \sum_{i=1}^n \lambda_d\big((y,x^{(n-1)})_i\big)  \rho_n(y,x^{(n-1)}) dx^{(n-1)} \\
	& =\sum_{n=2}^\infty n(n-1)  \int_{\mathbb{X}^{n-1}}	 \lambda_d(z) \rho_{n}(y,z,x^{(n-2)}) \,  dzdx^{(n-2)} \\
	& \qquad - \sum_{n=1}^\infty n \int_{\mathbb{X}^{n-1}}  \lambda_d(y)  \rho_n(y,x^{(n-1)}) dx^{(n-1)} - \sum_{n=1}^\infty n (n-1)\int_{\mathbb{X}^{n-1}}  \lambda_d(z)  \rho_n(y,z,x^{(n-2)}) dzdx^{(n-2)} \\
	& = -\sum_{n=1}^\infty  n\int_{\mathbb{X}^{n-1}}	 \lambda_d(y) \rho_{n}(y,x^{(n-1)}) \,  dx^{(n-1)}\\
	& = -\lambda_d(y) c(y),
\end{split}
\end{align}
where the two terms in the second and third line were handled as follows:
for the first one, we adjusted the index of the sum and for the second one, we split the inner sum into its first component and the rest. Gathering all the results above and reincorporating time dependence into the notation, we recover the PDE for the mean concentration
\begin{align}
\partial_t c(y,t) = D \nabla^2 c(y,t) -\lambda_d(y) c(y,t).
\end{align}

\subsection{Creation}

Analogously, for the spontaneous creation, \ce{$\emptyset$ ->A}, the CDME is
\begin{align}\begin{split} \label{eq:CDME_creation}
	\partial_t \rho_n(x^{(n)}) &=
	\sum_{i=1}^n D_i \rho_n (x^{(n)}) 
	- \int_{\mathbb{X}} \lambda_c(y)  \rho_n (x^{(n)}) \, dy \,
	+ \frac{1}{n}\sum_{i=1}^n \rho_{n-1}(x^{(n)}_{/i})  \lambda_c(x^{(n)}_i).
\end{split}\end{align}

Applying \cref{eq:conc} again, we can obtain an equation for the mean concentration. The diffusion will have the same form as in the example before, so we concentrate on the reaction terms,

\begin{align} \begin{split}
	\sum_{n=1}^\infty n &\int_{\mathbb{X}^{n-1}} \left( \frac{1}{n}\sum_{i=1}^n \lambda_c(x^{(n)}_i)\rho_{n-1}(x^{(n)}_{\setminus \{i\}})  - \int_{\mathbb{X}} \lambda_c(y)  \rho_n(x^{(n)}) \, dy \, \right) dx^{(n)}_{\setminus \{1\}} \\
	&= \sum_{n=1}^\infty  \int_{\mathbb{X}^{n-1}} \lambda_c(x^{(n)}_1)\rho_{n-1}(x^{(n)}_{\setminus \{1\}})dx^{(n)}_{\setminus \{1\}} +
	\sum_{n=1}^\infty (n-1) \int_{\mathbb{X}^{n-1}} \lambda_c(x^{(n)}_2)\rho_{n-1}(x^{(n)}_{\setminus \{2\}})dx^{(n)}_{\setminus \{1\}} \\
	& \qquad - \sum_{n=1}^\infty  n\int_{\mathbb{X}^{n}} \lambda_c(y)  \rho_n(x^{(n)}) \, dy dx^{(n)}_{\setminus \{1\}} \\
	&= \lambda_c(x^{(n)}_1) \sum_{n=1}^\infty  \int_{\mathbb{X}^{n-1}} \rho_{n-1}(x^{(n)}_{\setminus \{1\}})dx^{(n)}_{\setminus \{1\}} +
	\sum_{n=0}^\infty n \int_{\mathbb{X}^{n}} \lambda_c(x^{(n+1)}_{2})\rho_{n}(x^{(n+1)}_{\setminus \{2\}})dx^{(n+1)}_{\setminus \{1\}} \\
	& \qquad - \sum_{n=1}^\infty  n\int_{\mathbb{X}^{n}} \lambda_c(y)  \rho_n(x^{(n)}) \, dy dx^{(n)}_{\setminus \{1\}} \\
	&= \lambda_c(x^{(n)}_1).
\end{split}
\end{align}
In the second line, we split the first term into two parts, one for the first term of the inner sum and the other one for the rest. We later rearranged the index of the second term and used that the total probability sums to one in the first term. The PDE for the mean concentration reincorporating time in the notation is then 
\begin{align}
\partial_t c(y,t) = D \nabla^2 c(y,t) +\lambda_c(y).
\end{align}

\subsection{Mutual annihilation}

Let us consider the mutual annihilation reaction $A + A \rightarrow \emptyset$ with rate function $\lambda(x_1,x_2)$ depending on the reactant's positions. The corresponding CDME is given in \cref{eq:CDME_annihilation}, and it reads \cite{del2021probabilistic,delRazo2},

\begin{align}\begin{split} 
	\partial_t \rho_n(x^{(n)}) &=
	\mathcal{D}_n\rho_n(x^{(n)})  \quad + \frac{(n+1)(n+2)}{2} \int_{\mathbb{X}^2} \lambda(z_1,z_2)  \rho_{n+2}(z_1,z_2,x^{(n)}) \, dz_1 dz_2  \\
	& \quad  -\rho_n(x^{(n)}) \sum_{1\leq i <j \leq n}^n \lambda(x^{(n)}_i,x^{(n)}_j)   \,.
\end{split}
\end{align}

We apply \cref{eq:conc} again. The diffusion will have the same form as before, so we focus on the reaction terms. The calculations follow a similar logic as in the examples before. The reaction term is the sum of the gain term and the loss term. Applying \cref{eq:conc} to the gain term yields 

\begin{align} \begin{split} \label{eq:gain_mutannih}
	\sum_{n=1}^\infty n &\int_{\mathbb{X}^{n-1}} \left( \frac{(n+1)(n+2)}{2} \int_{\mathbb{X}^2} \lambda(z_1,z_2)  \rho_{n+2}(y,z_1,z_2,x^{(n-1)}) \, dz_1 dz_2  \right) dx^{(n-1)}\\
	&= 	\sum_{n=3}^\infty \frac{n(n-1)(n-2)}{2} \int_{\mathbb{X}^{n-1}} \lambda(z_1,z_2)  \rho_{n}(y,z_1,z_2,x^{(n-3)}) \, dz_1 dz_2 dx^{(n-3)},
\end{split}
\end{align}
where we simply adjusted the index of the sum and rewrote the integrals under one sign. Analogously, for the loss term
\begin{align}\begin{split} \label{eq:loss_mutannih}
	-\sum_{n=1}^\infty n &\int_{\mathbb{X}^{n-1}} \left( \rho_n(y,x^{(n-1)}) \sum_{1\leq i <j \leq n} \lambda\left( \left[y,x^{(n-1)}\right]_i,\left[y,x^{(n-1)}\right]_j\right)\right) dx^{(n-1)} \\
	&=-\sum_{n=1}^\infty n \int_{\mathbb{X}^{n-1}} \rho_n(y,x^{(n-1)})  \left( \sum_{j=2}^{n} \lambda\left(y,x^{(n-1)}_{j-1}\right) +  \sum_{2\leq i <j \leq n} \lambda\left( x^{(n-1)}_{i-1},x^{(n-1)}_{j-1}\right)\right) dx^{(n-1)} \\
	&=-\sum_{n=1}^\infty n (n-1) \int_{\mathbb{X}^{n-1}}  \rho_n(y,x^{(n-1)}) \lambda(y,x^{(n-1)}_1) dx^{(n-1)} \\
	& \qquad \qquad -\sum_{n=1}^\infty n \frac{(n-1)(n-2)}{2} \int_{\mathbb{X}^{n-1}}  \lambda(x^{(n-1)}_1,x^{(n-1)}_2) \rho_n(y,x^{(n-1)}) dx^{(n-1)} ,
\end{split}
\end{align}
where in the first step we separated the sum into the $i=1$ term and the rest. In the second step, we simply used the symmetry within the rate functions because particles of the same species are indistinguishable. We obtain the reaction term by summing up the gain and loss terms from \cref{eq:gain_mutannih} and \cref{eq:loss_mutannih}. Only one term survives:
\begin{align} \label{eq:term_second_order}
-\sum_{n=1}^\infty n (n-1) \int_{\mathbb{X}^{n-1}}  \rho_n(y,x^{(n-1)}) \lambda(y,x^{(n-1)}_1) dx^{(n-1)},
\end{align}
so the resulting equation is
\begin{align}
\partial_t c(y) = D \nabla^2 c(y) -\sum_{n=1}^\infty n (n-1) \int_{\mathbb{X}^{n-1}}  \rho_n(y,x^{(n-1)}) \lambda(y,x^{(n-1)}_1) dx^{(n-1)}.
\end{align}

This is not yet a closed equation, but it can be simplified further if we make additional assumptions on the form of the rate function. Analogous derivations for higher moments will be studied in other work that concentrates exclusively on calculating moment expansions of the CDME.

\section{Reaction schemes with space-dependent rates and reactive boundaries}
\label{sec:reactschemes}
Reactions can be implemented following well-known work on simulations of particle-based reaction-diffusion dynamics \cite{hoffmann2019readdy}. However, many of the well-known methods employ simple ad-hoc rate functions, without pointing a connection to an underlying master equation.  In the following, we overview and extend the schemes to simulate reactions in a spatially-dependent setting. Our approach is constructed for the arbitrary rate functions per reaction. Note we abuse the notation of $\lambda$ for rates functions, as it represents different functions for each case. In this section, we assume the reader is familiar with the well-known Gillespie \cite{gillespie1977exact,anderson2015stochastic} and $\tau$-leap algorithms\cite{anderson2015stochastic} for well-mixed reaction systems.

\subsection{Zeroth order reactions:}
These reactions involve no reactants; they simply add particles into the system. Thus, they do not depend on the state of the system, so it is possible to handle many reactions occurring in a given time-step. Let us first consider the simplest zeroth order reaction $\emptyset \rightarrow A$ with constant rate $\lambda$. At the density level, this corresponds to a creation reaction in the CDME \cref{eq:CDME_creat}. At the trajectory level, one needs to count the number of reactions $N(\tau)$ that occur in a time step $\tau$ is given in terms of the so-called random time-change Poisson representation \cite{anderson2015stochastic},
\begin{align}
	N(\tau) = Y\left(\lambda \tau\right),
	\label{eq:zerothRTC}
\end{align}
The  symbol $Y$ represents sampling from a unit rate Poisson process with a re-scaled time as argument \cite{anderson2015stochastic}. In other words $Y$ yields a random number of events---reactions--- that happen in time interval $\tau$ with rate $\lambda$. This is straightforward to implement computationally as shown in \cref{alg:reactions}. As each reaction implies adding one $A$, we add $N(\tau)$ $A$ particles uniformly in the simulation domain. In general, the rate can depend on time and space $\lambda(x,t)$. In such cases, we can generalize \cref{eq:zerothRTC} to
\begin{align}
	N(t+\tau) = Y\left(\int_{t}^{t+\tau}\Lambda(s) ds\right), \quad \text{where} \quad \Lambda(t)=\int_\mathbb{X}\lambda(x,t)dx,
	\label{eq:zerothRTC2}
\end{align}
and $\mathbb{X}$ represent the full spatial domain of the system. \Cref{eq:zerothRTC2} can be solved exactly with the Gillespie algorithm, also called stochastic simulation algorithm \cite{anderson2015stochastic,gillespie1977exact}. It is however, more computationally efficient to approximate it using the $\tau-$leap algorithm \cite{anderson2015stochastic},
\begin{align}
	N(t+\tau) = Y\left(\Lambda(t) \tau\right),
	\label{eq:zerothRTC3}
\end{align}
where the time-step $\tau$ is small. How small it must chosen will depend on the specific system and rate function. This can be implemented computationally as shown in \cref{alg:reactions}. Note the $\tau-$leap approximation consists on assuming the rate $\Lambda$ is constant in time along the time-step $\tau$. \Cref{eq:zerothRTC2} yields the number of particles one needs to add into the system, but we still don't know where. This can be simply solved by normalizing the rate function into a probability density
\begin{align}
	\rho(x,t) = \frac{\lambda(x,t)}{\int_{\mathbb{X}}\lambda(x,t) dx},
	\label{eq:distSampling}
\end{align}
then we sample the positions of $N(t+\tau)$ particles from this distribution, for instance with inverse transform sampling. Finally, we simply add the particles at the sampled positions. 

In the case of reactions with more than one product, we can apply an analogous methodology. For instance, let us consider the reaction $\emptyset \rightarrow A+B$ with rate function $\lambda(x,y,t)$, where $x$ and $y$ correspond to the positions of the products $A$ and $B$,  respectively. To sample the number of reactions, we again use \cref{eq:zerothRTC3} with 
$\Lambda(t)=\int_{\mathbb{X}^2}\lambda(x,y,t) dx dy$. Then, we sample their positions from the distribution
\begin{align}
	\rho(x,y,t) = \frac{\lambda(x,y,t)}{\int_{\mathbb{X}^2}\lambda(x,y,t) dx dy}.
\end{align}
Finally, we add the particles at the sampled positions.

\subsection{First order reactions:}
These reactions involve only one reactant, for example, $A\rightarrow \emptyset$. Unlike zeroth order reactions, first order reactions do depend on the state of the system, so it is not possible to handle many reactions in one time step. Thus we must assume that the integration time-step is small enough such that the products of a reaction that occurred in a given time step are not involved in another reaction within the same time step. In other words, we assume the integration time-step is significantly smaller than the inverse of the largest (non-zeroth order) reaction rate of a given system \cite{hoffmann2019readdy}.

Let us first assume reactions occur continuously and independently at a constant average rate $\lambda$. This means that the time between reaction events is exponentially distributed, and thus the probability of a reaction occurring within a time interval $\tau$ is given by
\begin{align}
	p(\lambda, \tau) = 1 - e^{-\lambda \tau}.
	\label{eq:probReact}
\end{align}
To calculate if a reaction occurred, we sample a random number at each time step of length $\tau$ and check if it is below $p(\lambda, \tau)$. If so, then the reaction occurs, and we add or remove particles/molecules accordingly. If the rate is time and/or position dependent $\lambda(t,x(t)$, one can generalize the calculation 
\begin{align}
	p\left(\lambda(t,x(t)\right), \tau) = 1 - \exp{\left(-\int_t^{t+\tau} \lambda\left(s,x(s)\right)ds\right)} \approx 1 - e^{-\lambda(t,x(t))\tau},
	\label{eq:probReact2}
\end{align}
where the first order approximation holds for small time steps $\tau$, and one can use this expression to derive higher-order implicit approximations. The pseudo-code implementation is shown in \cref{alg:reactions}.

Note this method can also sample zeroth order reactions. However, it will limit the amount of reactions to one per time-step. This will be accurate if the integration time-step is considerably smaller than the inverse of the corresponding zeroth order reaction rate. If we have a reaction with products, such as $A\rightarrow 2B$, we can apply a similar approach to that of zeroth order reactions. We will also explore this case in more detail within the next subsection.

\subsection{Second order reactions:}
\label{sec:secordreactSims}

\begin{figure}
	\begin{minipage}{0.49\linewidth}
		\begin{algorithm}[H]
			\DontPrintSemicolon
			\SetKwProg{Fn}{Function}{:}{}
			\SetKwFunction{FDiffusion}{Diffusion}
			\Fn{\FDiffusion{$\tau$}}{
				$x(t+\tau) \gets x(t) + \sqrt{2D\tau} \mathcal{N}(0,1)$
			}
			\vspace{2mm} 
			\SetKwFunction{FZerothReactions}{ZerothOrderReactions}
			\Fn{\FZerothReactions{$\tau, \lambda(x), \Lambda$}}{
				$N \gets$  Poisson($\Lambda \tau$) \\
				$\rho(x) \gets \lambda(x)/\Lambda$ \\
				\For{$j=1$ to $N$}{
					x $\gets$ Sample from $\rho(x)$ \\
					add particle at x
				}
			}
			\vspace{2mm}
			\SetKwFunction{FUniReactions}{UnimolecularReactions}
			\Fn{\FUniReactions{$\tau,\lambda(x)$}}{
				\For{particle in particleslist}{
					x $\gets$ particle.position\\
					\textit{react-prob} $\gets 1-\exp(-\lambda(\mathrm{x}) \tau)$ \\
					$r \gets$ random($0,1$) \\
					\If{$r<=$ react-prob}{
						reaction happens (add/remove particles)
					}
				}
			}
			\vspace{2mm}
		\end{algorithm}
	\end{minipage}
	\begin{minipage}{0.49\linewidth}
		\begin{algorithm}[H]
			\DontPrintSemicolon
			\SetKwProg{Fn}{Function}{:}{}
			\SetKwFunction{FBiReactions}{BimolecularReactions}
			\Fn{\FBiReactions{$\tau,\lambda(x,y)$}}{
				\For{$p_i,p_j$ in all possible pairs of particleslist}{
					$\mathrm{x}_i, \mathrm{x}_j \gets$ $p_i$.position, $p_j$.position\\
					\textit{react-prob} $\gets 1-\exp(-\lambda(\mathrm{x}_i,\mathrm{x}_j) \tau)$ \\
					$r \gets$ random($0,1$) \\
					\If{$r<=$ react-prob}{
						reaction happens (add/remove particles)
					}
				}
			}
			\vspace{2mm}
			\SetKwFunction{FBiReactions}{BimolecularReactionsDoi}
			\Fn{\FBiReactions{$\tau,\kappa,\sigma$}}{
				\For{$p_i,p_j$ in all possible pairs of particleslist}{
					$\mathrm{x}_i, \mathrm{x}_j \gets$ $p_i$.position, $p_j$.position\\
					\If{$|\mathrm{x}_i-\mathrm{x}_j|\leq \sigma$}{
						\textit{react-prob} $\gets 1-\exp(-\kappa \tau)$ \\
						$r \gets$ random($0,1$) \\
						\If{$r<=$ react-prob}{
							reaction happens (add/remove particles)
						}
					}
				}
			}
			
			\vspace{2mm}
		\end{algorithm}
	\end{minipage}
	\caption{Pseudo-code implementation of diffusion and reactions procedures for one time step $\tau$. All the rates in the functions arguments are assumed to be evaluated at the current time $t$. The first routine correspond to the Euler-Maruyama scheme for the diffusion of particles following standard Brownian motion (overdamped Langevin dynamics). The second routines handles	zeroth order reaction with arbitrary rate functions using $\tau$-leaping, where Poisson($\Lambda \tau$) means sampling from a unit rate Poisson process for a rescaled time interval of $\Lambda \tau$. The third procedure handles unimolecular reactions. The fourth and fifth procedures handle bimolecular reactions with either an arbitrary rate function or with the Doi approach (\cref{eq:DoiRate}), respectively. If particles need to be added, their location is sampled the same way as for zeroth order reactions.
	} 
	\label{alg:reactions}
\end{figure}

Second order reactions involve two reactants, so they are also called bimolecular reactions. Assuming there are no products, i,e, $A+B\rightarrow \emptyset$, the rate function is $\lambda(x,y)$, and it depends on the reactants positions $x$ and $y$. As the positions themselves are functions of time, one can calculate the probability of a reaction occurring by calculating  $p(\lambda\left(x(t),y(t)\right),\tau)$ in \cref{eq:probReact2},
\begin{align}
	p(\lambda(x(t),y(t)),\tau)\approx 1 - e^{-\lambda(x(t),y(t))\tau}.
	\label{eq:probReactBim}
\end{align}
Time dependence can also be included analogously to \cref{eq:probReact2}. One can often safely assume that $\lambda$ depends only on the relative distance between the reactants $|x-y|$ and that it decays fast as this distance grows. In this case, one can approximate the rate function by a discretized rate function
\begin{align}
	\lambda(x,y)= \kappa \ \mathds{1}_{|x-y|\leq \sigma},
	\label{eq:DoiRate}
\end{align}
where $\mathds{1}_\omega$ is the indicator function on set $\omega$ and $\sigma$ is chosen as an interaction cutoff. The constant $\kappa$ must be chosen so the integral over the relative distance of the original and the discretized rate functions match, i.e.
\begin{align}
	\kappa \ = \frac{1}{V_\sigma}\int_\mathbb{X} \lambda(x_0,y)dy,
\end{align}
where ${V_\sigma}=\int_\mathbb{X}\mathds{1}_{|x_0-y|\leq \sigma}dy $ is the volume of the reactive region that corresponds to $4\pi \sigma^3/3$ in three dimensions, $x_0$ is arbitrary since $\lambda$ only depends on the relative distance and $\mathbb{X}$ is the whole spatial domain. This discretization corresponds to the volume reactivity model of Doi\cite{doi1976stochastic}, where particles $A$ and $B$ react with rate $\kappa$ only if they are within a distance $\sigma$ of each other. This convention is the one used in many simulations and software packages \cite{hoffmann2019readdy}. To implement in simulations, one first checks if particles are within a distance $\sigma$ of each other. If so, then we calculate the probability of reaction with $p(\kappa,\tau)$ in \cref{eq:probReact} and proceed analogously as before. 

A more general case is for the reaction $A+B \rightarrow C$ as shown in the CDME from \cref{sec:cdme_bimol}. In this case, the rate function is $\lambda(z;x,y)$, where $x$ and $y$ are the positions of the reactants and $z$ is the location of the product. We want to once again check if a reaction occurs. As we don't know where the products will be located, we calculate this regardless of where the products will be located, so we integrate over the product position $z$ 
\begin{align}
	\tilde{\lambda}(x,y) = \int_\mathbb{X} \lambda(z;x,y) dz.
\end{align}
Note this integration also appears in the CDME loss term (\cref{eq:CDMEbim}). We then proceed in the same manner as before by calculating $p(\tilde{\lambda}(x(t),y(t)),\tau)$ using \cref{eq:probReactBim}. If a reaction occurs, we still need to sample the position of the product $C$. For a reaction that occurred at $x_0$ and $y_0$, the product can then be sampled from the distribution 
\begin{align}
	\rho(z,t) = \frac{\lambda(z;x_0,y_0)}{\int_{\mathbb{X}}\lambda(z;x_0,y_0) dz},
\end{align}
analogously to \cref{eq:distSampling}. A common choice in software packages \cite{hoffmann2019readdy} is to place $C$ always in the middle distance between the reactants, i.e.  $\rho(z;x_0,y_0)=\delta(z-(x_0+y_0)/2)$. Combining this with \cref{eq:DoiRate}, we can explicitly write the rate function for the CDME that corresponds to the schemes often used in software packages \cite{hoffmann2019readdy} as
\begin{align}
	\lambda(z;x,y) = \kappa \ \mathds{1}_{|x-y|\leq \sigma} \ \delta\left(z-\frac{x+y}{2}\right).
\end{align}
This provides a direct connection between existing simulation schemes \cite{hoffmann2019readdy} and the underlying CDME for the probabilistic dynamics. Unlike most methods often used in software packages, the schemes developed here are for a general rate function and thus not limited to this specific form. Note when considering reversible reactions, from a physics perspective, one might need to take into account detailed balance when determining the rate functions of the reverse reaction \cite{frohner2018reversible}. The relevant notable work \cite{isaacson2018unstructured} also employs general rate functions and takes into account detailed balance for reversible bimolecular reactions in the context of reaction-diffusion master equations. Also note that as we are at the particle level, to model and simulate reactions involving three or more reactants (higher-order), we can always handle them as a combination of two or more second order reactions.

\subsection{Reactive boundaries: absorbing and partially absorbing boundaries}
\label{sec:PRBM}

\begin{figure}
	\begin{minipage}{0.49\linewidth}
		\begin{algorithm}[H]
			\DontPrintSemicolon
			\SetKwProg{Fn}{Function}{:}{}
			\SetKwFunction{FPABoundary}{PartiallyAbsorbingBoundary}
			\Fn{\FPABoundary{$\tau,\tilde{\lambda}_r$}}{
				\textit{react-prob} $\gets 1-\mathrm{exp}(-\tilde{\lambda}_r \tau) $ \\
				\For{particle in particleslist}{
					\If{particle.position $\in \Sigma$}{
						$r \gets$ random($0,1$) \\
						\If{$r<=$ react-prob}{
							remove particle
						}
					}
				}
			}
			\vspace{2mm}
		\end{algorithm}
	\end{minipage}
	\begin{minipage}{0.49\linewidth}
		\begin{algorithm}[H]
			\DontPrintSemicolon
			\SetKwProg{Fn}{Function}{:}{}
			\SetKwFunction{FABoundary}{AbsorbingBoundary}
			\Fn{\FABoundary{}}{
				\For{particle in particleslist}{
					\If{particle.position $\in \Sigma$}{
						remove particle
					}
				}
			}
			\vspace{2mm}
			\SetKwFunction{FABoundaryalt}{AbsorbingBoundary2}
			\Fn{\FABoundaryalt{}}{
				\For{particle in particleslist}{
					\If{particle.position $\leq \sigma$}{
						remove particle
					}
				}
			}
			\vspace{2mm}
		\end{algorithm}
	\end{minipage}
	\caption{Pseudo-code implementation of  partially absorbing boundary and absorbing boundary for one time step $\tau$. We also show two possible implementations of the absorbing boundary: the first one correspond to the infinite reaction rate limit of the partially absorbing one; the second one corresponds to an absorbing boundary condition for the diffusion. They are equivalent for infinitesimal small boundary layer $\Sigma$. 
	} 
	\label{alg:absorbingBound}
\end{figure}

Partially absorbing boundaries are often used in diffusion-influenced reaction models \cite{collins1949diffusion,doi1976stochastic,shoup1982role}. At the macroscopic level, they correspond to the first boundary condition from \cref{eq:smoleqBC}, i.e. material flux across the boundary proportional to the concentration at the boundary. At the particle level, they can be understood as partially reflected Brownian motion \cite{del2016discrete, grebenkov2006partially}. This means Brownian particles can be either reflected from a boundary or absorbed by it; at which degree one or the other happens depends on some given rate.  Numerical schemes for partially reflected Brownian motion can be implemented using the methods presented in \cite{del2016discrete}. 

In this work, we present a more straightforward approach based on the CDME. In \cref{sec:diffinflCDMElimit} we showed that one can recover the partially absorbing boundary condition (\cref{eq:smoleqBC}) as a limiting case of the CDME. Thus, we can obtain a numerical scheme to model the partially absorbing boundary by discretizing the rate $\lambda_r$ from \cref{eq:diffinflCDMErates}. We could discretize the Dirac delta using \cref{eq:discretedeltas}. However, the most straightforward discretization is to use an indicator function along a boundary layer $\Sigma$ of length $\Delta r$, $\Sigma =\{r|\sigma \leq r\leq \sigma+\Delta r\}$, which yields
\begin{align}
	\tilde{\lambda}_r = \frac{\kappa}{V_\Sigma} \mathds{1}_{x\in\Sigma}, 
	\label{eq:disclamr}
\end{align}
where $V_\Sigma$ is the volume of the boundary layer $\Sigma$. As we are assuming a spherical domain, the boundary layer corresponds to a spherical shell and thus
\begin{align}
	V_\Sigma=\frac{4\pi}{3}[(\sigma+\Delta r)^3- \sigma^3].
\end{align}
Note if we integrate \cref{eq:disclamr} in the whole three-dimensional domain, we obtain simply $\kappa$, which is the same as integrating $\lambda_r$ from \cref{eq:diffinflCDMErates}, so the discretization is consistent. Moreover, the first order term corresponds to $V_\Sigma\approx 4\pi \sigma^2 \Delta r$, so this rate is also consistent with the results in \cite{del2016discrete}. Alternative discretizations of $\tilde{\lambda}_r$ are also possible, and the following procedure applies analogously.

To implement reactions due to a partially absorbing boundary, one needs to sample the possibility of a reaction event with rate $\tilde{\lambda_r}$ within one time step $\tau$ for every particle in $\Sigma$ , i.e 
\begin{align*}
	p(\tilde{\lambda}_r, \tau) = 1-e^{-\tilde{\lambda}_r \tau}.
\end{align*}
If the reaction happens, the particle is removed. The implementation is shown in the \textit{PartiallyAbsorbingBoundary} function in \cref{alg:absorbingBound}. For completeness, we also show two possible implementations for the fully absorbing boundary case.

\bibliographystyle{abbrvurl}
\bibliography{references}

\end{document}